\documentclass[%
 superscriptaddress,
preprint,
 amsmath,amssymb,
 aps,
]{revtex4-2}

\usepackage[color=yellow]{todonotes} 
\newcommand*\chem[1]{\ensuremath{\mathrm{#1}}}
\usepackage{siunitx}
\usepackage{graphicx}
\usepackage{dcolumn}
\usepackage{bm}
\usepackage{xcolor}
\usepackage{comment}
\usepackage{float}
\usepackage{tablefootnote}

\usepackage{subfig}
\begin{document}

\preprint{AIP/123-QED}

\title{Ultrasonic Study of Water Adsorbed in Nanoporous Glasses}

\author{Jason Ogbebor}
\affiliation{Otto H. York Department of Chemical and Materials Engineering,\\
New Jersey Institute of Technology,\\
323 Dr. Martin Luther King Jr. Blvd, Newark, NJ 07102, USA}
\author{John J. Valenza}
\affiliation{ExxonMobil Technology and Engineering Co., Research Division, 1545 Route 22 East, Annandale, NJ, 08801, USA}
\author{Peter I. Ravikovitch}
\affiliation{ExxonMobil Technology and Engineering Co., Research Division, 1545 Route 22 East, Annandale, NJ, 08801, USA}
\author{Ashoka Karunarathne}
\affiliation{Otto H. York Department of Chemical and Materials Engineering,\\
New Jersey Institute of Technology,\\
323 Dr. Martin Luther King Jr. Blvd, Newark, NJ 07102, USA}
\author{Giovanni Muraro}
\affiliation{ExxonMobil Technology and Engineering Co., Research Division, 1545 Route 22 East, Annandale, NJ, 08801, USA}
\author{Maxim Lebedev}
\affiliation{Center for Exploration Geophysics, Curtin University, 26 Dick Perry Avenue, Kensington, 6151, WA, Australia}
\affiliation{Centre for Sustainable Energy and Resources,  Edith Cowan University, 270 Joondalup Drive Joondalup, 6027, WA, Australia}
\author{Boris Gurevich}
\affiliation{Center for Exploration Geophysics, Curtin University, 26 Dick Perry Avenue, Kensington, 6151, WA, Australia}
\author{Alexei F. Khalizov}
\affiliation{Department of Chemistry and Environmental Science,\\
New Jersey Institute of Technology,\\
323 Dr. Martin Luther King Jr. Blvd, Newark, NJ 07102, USA}
\affiliation{Otto H. York Department of Chemical and Materials Engineering,\\
New Jersey Institute of Technology,\\
323 Dr. Martin Luther King Jr. Blvd, Newark, NJ 07102, USA}
\author{Gennady Y. Gor}
\email{E-mail: gor@njit.edu}
\affiliation{Otto H. York Department of Chemical and Materials Engineering,\\
New Jersey Institute of Technology,\\
323 Dr. Martin Luther King Jr. Blvd, Newark, NJ 07102, USA}


\begin{abstract}
Thermodynamic properties of fluids confined in nanopores differ from those observed in the bulk. To investigate the effect of nanoconfinement on water compressibility, we performed water sorption experiments on two nanoporous glass samples while concomitantly measuring the speed of longitudinal and shear ultrasonic waves in these samples. These measurements yield the longitudinal and shear moduli of the water laden nanoporous glass as a function of relative humidity that we utilized in the Gassmann theory to infer the bulk modulus of the confined water. This analysis shows  that the bulk modulus (inverse of compressibility) of confined water is noticeably higher than that of the bulk water at the same temperature. Moreover, the modulus exhibits a linear dependence on the Laplace pressure. The results for water, which is a polar fluid, agree with previous experimental and numerical data reported for non-polar fluids. This similarity suggests that irrespective of intermolecular forces, confined fluids are stiffer than bulk fluids. Accounting for fluid stiffening in nanopores may be important for accurate interpretation of wave propagation measurements in fluid-filled nanoporous media, including in petrophysics, catalysis, and other applications, such as in porous materials characterization.

\end{abstract}

\maketitle

\section{Introduction}

Since the time of Lord Kelvin~\cite{Kelvin1872} and Derjaguin~\cite{Karasev1971,Derjaguin1986}, numerous works have demonstrated that confinement affects fluid behavior, such as: evaporation/condensation and freezing/melting phase transitions~\cite{Huber2015, Gubbins2014, Barsotti2016}, transport properties~\cite{Kimmich2002, Xu2009transport}, and derivative thermodynamic properties, such as the thermal expansion coefficient~\cite{Xu2009, Valenza2005} and compressibility~\cite{Schappert2014, Dobrzanski2021}.

Over the past two decades an additional motivation for studying confined fluids emerged from the shale revolution. Extended reach horizontal drilling and hydraulic fracturing unlocked the potential of unconventional oil and gas assets commonly referred to as oil and gas shale. This industrial paradigm shift motivated interest in developing petrophysical techniques to characterize these assets. Conventional petrophysics is primarily concerned with mechanisms active on a length scale greater than one micron~\cite{Muller2005}. In contrast, unconventional oil and gas reservoirs primarily consist of oil and gas confined in nanoporous mudstones ~\cite{Valenza2013}. 

The seminal works of Gassmann~\cite{Gassmann1951} and Biot~\cite{Biot1956i,Biot1956ii} showed how the elastic properties (compressibility, or, the reciprocal property, bulk modulus) of a pore fluid contribute to the effective mechanical and elastodynamic response of a saturated porous medium. A number of petrophysical sonic logging techniques exploit this poromechanical response to infer the pore fluid properties. Elastic properties of a porous medium (be it dry or saturated with a fluid) can be inferred by measuring the speed of various stress waves. Utilizing several fluids, previous experimental studies explored the relationship between sound speed and relative saturation of nanoporous media \cite{Pimienta2019, Tadavani2020}. All previous work indicates that fluids confined to nanopores exhibit elastic properties that deviate from those in the bulk~\cite{Warner1988, Page1995, Schappert2013JoP, Schappert2014, Schappert2015EPL, Schappert2018, Schappert2018liquid, Schappert2022}. Recent density functional theory calculations~\cite{Gor2014, Sun2019density} and molecular simulations~\cite{Dobrzanski2018, Gor2018Gassmann, Maximov2018} confirm these observations.

While natural geological materials are of primary practical interest, these materials are characterized by a great deal of complexity such as a broad pore size distribution, and the presence of multiple components with disparate chemical makeup. To understand how confinement affects the elasticity of the fluid, it is more convenient to use a simpler porous medium with a single solid phase and a microstructure characterized by a large surface area and narrow pore size distribution. Since we use sound to probe the mechanical properties, it is also preferable to work with a monolithic porous sample as opposed to granular media, which are known to exhibit large sound speed variations as a function of sample preparation (e.g., leading to different packing densities between samples or even inhomogeneities within the same sample). Vycor glass is a suitable model nanoporous medium that adheres to the constraints listed above, and the solid skeleton is silica, a primary component in many rocks, including mudstones. Moreover Vycor has been utilized in the adsorption-ultrasonic studies mentioned above, focused on confined liquid nitrogen~\cite{Warner1988, Schappert2022}, argon~\cite{Schappert2013JoP, Schappert2014, Schappert2018, Schappert2018liquid} and linear alkanes~\cite{Page1995, Schappert2015EPL}. In addition, there are numerous experiments concerned with water confined in Vycor glass that explored adsorption-induced deformation~\cite{Amberg1952, Hiller1964}, freezing~\cite{Hodgson1960, Petrov2011}, imbibition~\cite{Gruener2012, Kiepsch2016, Gruener2016PRE}, {the temperature of maximum density~\cite{Taschin2010},} and the structure and fluid dynamics of water in the pores~\cite{Bellissent2001, Thompson2007, Borman2016, Levitz2019}. Note that Murphy~\cite{Murphy1982} carried out an adsorption-acoustic study of water in Vycor glass, but focused on the attenuation of low-frequency acoustic waves, and did not probe the elasticity of water in the pores. {Cucini et al. also studied sound propagation in water-saturated Vycor glass, focusing on the dynamics of water in  nanopores, but not the elasticity~\cite{Cucini2010}. Thus, there are no experiments focused on the effect of nanoporous confinement on the compressibility of water via adsorption-acoustic methods.} Here we present the first experimental study of water vapor adsorption on nanoporous glass samples accompanied with \textit{in situ} ultrasonic measurements, probing the elasticity of water-laden porous glass. Contrary to similar previous experimental work~\cite{Warner1988, Page1995, Schappert2013JoP, Schappert2014, Schappert2018, Schappert2018liquid, Schappert2022} that utilize non-polar fluids, water is polar so it may interact differently with the glass surface. Furthermore, water is a substance ubiquitous in nature and is relevant to the geological context discussed previously.

We performed water sorption experiments on two nanoporous glass samples while concomitantly measuring the speed of longitudinal and shear ultrasonic waves in these samples. This includes both the adsorption and desorption branches of the isotherm. Using these measurements, we calculated the longitudinal and shear moduli of the samples as a function of relative humidity (RH). These moduli are then used to compute the bulk modulus, which shows a sharp increase for both samples when the pores become completely filled with the liquid water. These elevated bulk moduli of fully saturated samples are then used to infer the bulk modulus of the confined water using the Gassmann theory. Our observations also show a significant hysteresis effect, in which the moduli remain elevated until capillary evaporation of the pore fluid occurs. The bulk modulus of confined water estimated with the Gassmann model is higher than the modulus of the bulk water obtained from an equation of state. Our observations are consistent with previous experimental and computational studies on non-polar liquids, which suggests that confinement enhances liquid modulus irrespective of molecular characteristics like polarizability.

\section{Methods}

\begin{figure}
\centering
\includegraphics[width=1.0\linewidth]{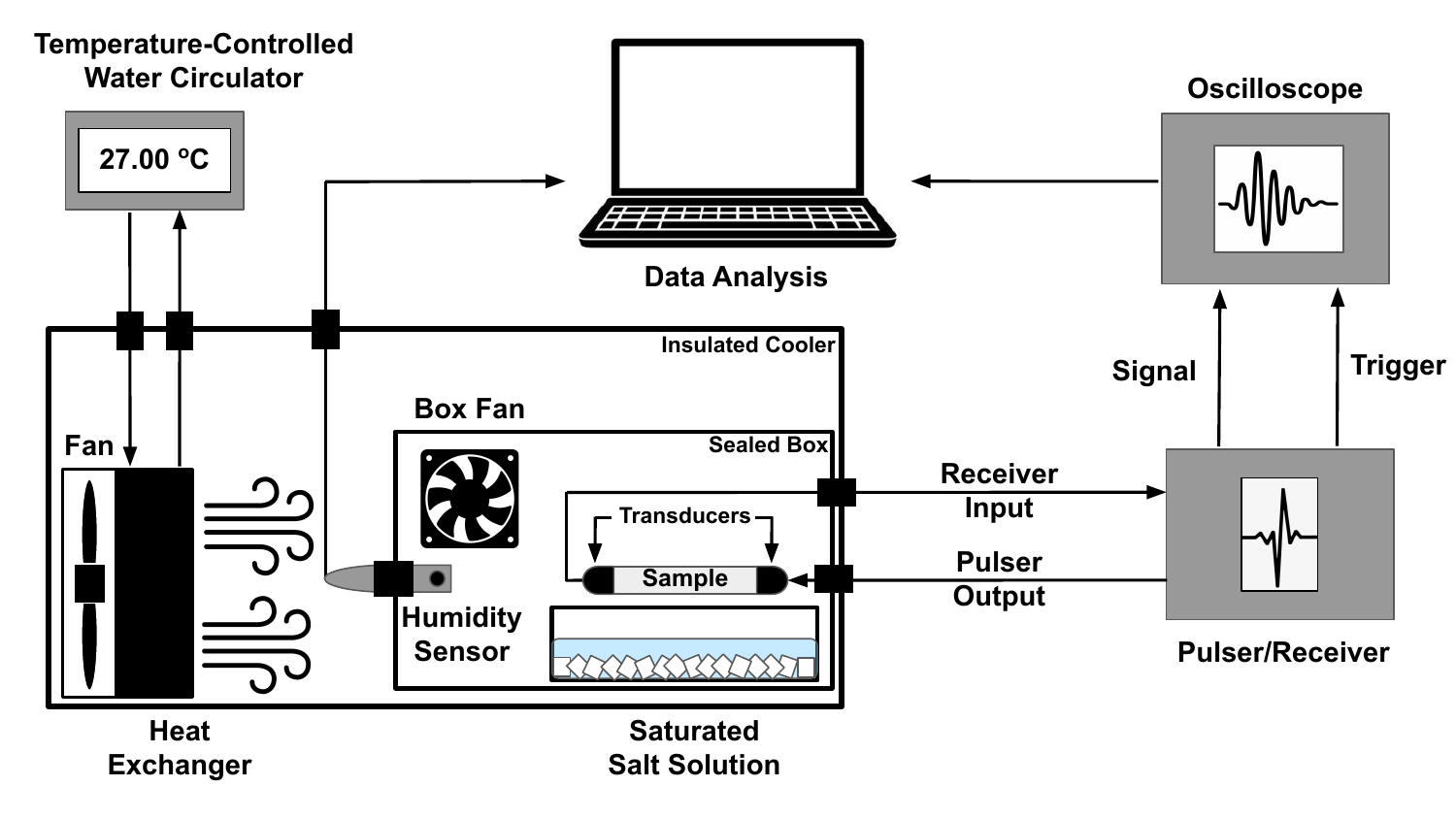}
\caption{Schematic of the adsorption-ultrasound system. The cables for the transducers, humidity sensor, and box fan were fed into the chamber using sealed feed-throughs. Thermal insulation was applied to seal the holes through which cables/tubes passed through the walls of the cooler.}
\label{fig:Schematic}
\end{figure}

In general, our approach consists of measuring the speed of stress waves through a nanoporous medium as the porosity is gradually filled with liquid water. The water saturation is controlled by varying the {relative humidity} around the two {nanoporous} samples. {Since we can not measure the time of flight and weight of the sample at the same time,} after the RH equilibrates, the {time} of flight is recorded on one sample, and the other sample is weighed, then the RH is adjusted to a new value. These steps are repeated until both the adsorption and desorption isotherms are complete. {Then} the samples are swapped and the entire set of measurements is repeated. A schematic of the adsorption-ultrasound system is shown in Figure~\ref{fig:Schematic}, and the measurements are described in greater detail below.

\subsection{Porous Glass}

This study was conducted on two nanoporous glass rods of similar composition and pore structure, which will be referred to as samples ``V" and ``C". Sample V was Corning Vycor glass (No. 7930) with diameter $d_{\rm V}=$ 0.635 cm and length $L_{\rm V}=$ 2.540 cm. Vycor is an optically transparent porous glass, which is mostly silica. Its structure is comprised of worm-like channel pores with a mean diameter of 7-8 nm. Due to its accessible and well-defined pore space, optical transparency, and ability to be manufactured in a variety of shapes, Vycor has been utilized as a model porous medium for studying the physics of confined phases for several decades~\cite{Amberg1952, Murphy1982, Warner1988, Page1995, Thompson2007, Petrov2011, Gruener2012, Kiepsch2016, Schappert2014, Schappert2018,  Borman2016, Levitz2019, Schappert2022}. Sample C was a controlled-pore glass with diameter $d_{\rm C}=$ 0.622 cm and length $L_{\rm C}=$ 3.544 cm. The physical properties of both samples are listed in Table~\ref{tab:Samples}. All dimensional measurements have an uncertainty of $\pm$ 0.002 cm. The apparent density is the mass of the dry sample divided by the measured volume, which includes pore volume. The solid density is calculated as the dry mass divided by the solid volume [measured volume multiplied by $(1-\phi)$].

Both samples were characterized by nitrogen adsorption porosimetry to determine the porosity and pore size distributions. Adsorption isotherms of nitrogen at 77~K were measured with the volumetric adsorption instrument Autosorb-1C (Anton Paar). {The samples were out-gassed using a turbomolecular pump to residual pressure below \SI{7.5E-4}{torr}} at \SI{300}{\degree C} for 16 hours prior to running the isotherms. The entire monolithic glass rods (with mass exceeding 1~g in both cases) were used for the measurements. Due to the large size, the measurements took four days, over which the Dewar was carefully refilled with nitrogen when the pressure was greater than that corresponding to capillary condensation. The pore-size distributions (PSD) was calculated using the non-local density functional theory (NLDFT) model for cylindrical pores. The adsorption branch was used to determine the PSD along with the model for metastable adsorption, as in Ref.~\onlinecite{Ravikovitch2001colloids}. The PSD of Vycor glass is in good agreement with previously reported measurements \cite{Thommes2006}. 

{The samples used in this study were exposed to ambient air before and during the experiments detailed herein. At the end of adsorption-ultrasonic experiments the samples were vacuumed and heated. As a result of heating the samples became yellow in color (Figure~\ref{fig:Sample_Holder}). The color was likely caused by products of thermal decomposition of trace organic molecules adsorbed from air.}

\begin{table*}
\caption{Physical properties of sample V (Vycor glass) and C (Controlled-pore glass).}
\label{tab:Samples}
\begin{ruledtabular}
\begin{tabular}{lcc}
Property & Sample V & Sample C \\
\colrule
Dry mass (g) & 1.1363 $\pm$ 0.0001 & 1.6161 $\pm$ 0.0001 \\
Porosity, $\phi$ & 0.330 $\pm$ 0.001 & 0.332 $\pm$ 0.001 \\
Apparent density ($\rm g/cm^3$) & 1.412 $\pm$ 0.009 & 1.486 $\pm$ 0.009 \\
Solid density ($\rm g/cm^3$) & 2.108 $\pm$ 0.007 & 2.247 $\pm$ 0.008 \\
Mean pore diameter (nm) & 7.5 & 7.2 \\
Specific surface area ($\rm m^2/g$) & 145 & 158 \\
Specific pore volume ($\rm cm^3/g$) & 0.220 & 0.225 \\
\end{tabular}
\end{ruledtabular}
\end{table*}

\subsection{Humidity Control}

Water sorption experiments were carried out in humid air. The RH $= p/p_0$, is the ratio of the partial vapor pressure, $p$, to the saturation vapor pressure of water, $p_0$. The RH is controlled by placing various salt solutions (following Yurikov et al.~\cite{Yurikov2018}) \ in a hermetically sealed 2.5 L anaerobic box (AnaeroPack$^{\rm TM}$ rectangular jar, Thermo Scientific$^{\rm TM}$ R681001). Each solution contained enough excess salt to yield precipitate so the water activity remained constant. A list of salt solutions and the corresponding RH at room temperature is shown in Table~\ref{tab:Salts}. We added multiple wire feed-throughs to the box that permitted continuous monitoring of the relative humidity in the chamber using a Vernier$^{\rm TM}$ sensor with accuracy of $\pm$2\% RH, a resolution of 0.01\% RH, and a response time (time for a 90\% change in reading) of $\sim$40 seconds in well-mixed air. A 40 mm, 12V DC box fan continuously mixed the air in the box. Finally the box was placed in a Coleman 48 quart beverage cooler equipped with a finned shell-and-tube heat exchanger and a 120 mm 24V DC box fan. Temperature-controlled water was pumped through the heat-exchanger by a Fisherbrand$^{\rm TM}$ Isotemp$^{\rm TM}$ bath circulator, model 6200 R20, with the temperature set to $27.00 \pm 0.025$ $^{\circ}$C.

\begin{table}[b]
\caption{Desiccant salts used in this work, and their expected relative humidities at $T = \SI{27}{\degree C}$~\cite{CRC_Handbook}.}
\label{tab:Salts}
\begin{ruledtabular}
\begin{tabular}{ccc}\\
Salt name & Formula & RH (\%) \\
\colrule
Lithium chloride & \chem{LiCl} & 12 \\
Potassium acetate\footnotemark[1] & \chem{KC_2H_3O_2} & 24 \\
Calcium chloride & \chem{CaCl_2} & 30 \\
Potassium carbonate & \chem{K_2CO_3} & 45 \\
Magnesium nitrate & \chem{Mg{(NO_3)}_2} & 53 \\
Sodium bromide & \chem{NaBr} & 58 \\
Sodium chlorate & \chem{NaClO_3} & 75 \\
Sodium chloride & \chem{NaCl} & 77 \\
Ammonium sulfate & \chem{(NH_4)_2SO_4} & 81 \\
Potassium bromide & \chem{KBr} & 84 \\
Potassium bisulfate & \chem{KHSO_4} & 86 \\
Potassium chloride & \chem{KCl} & 88 \\
Sodium sulfate & \chem{Na_2SO_4} & 93 \\
Potassium sulfate & \chem{K_2SO_4} & 98 \\
\end{tabular}
\end{ruledtabular}
\footnotetext[1]{The data for potassium acetate is not included in the results, it is addressed in the Appendix.}
\end{table}

\begin{figure}
\centering
\subfloat[Sample V]{ \includegraphics[width=0.48\linewidth]{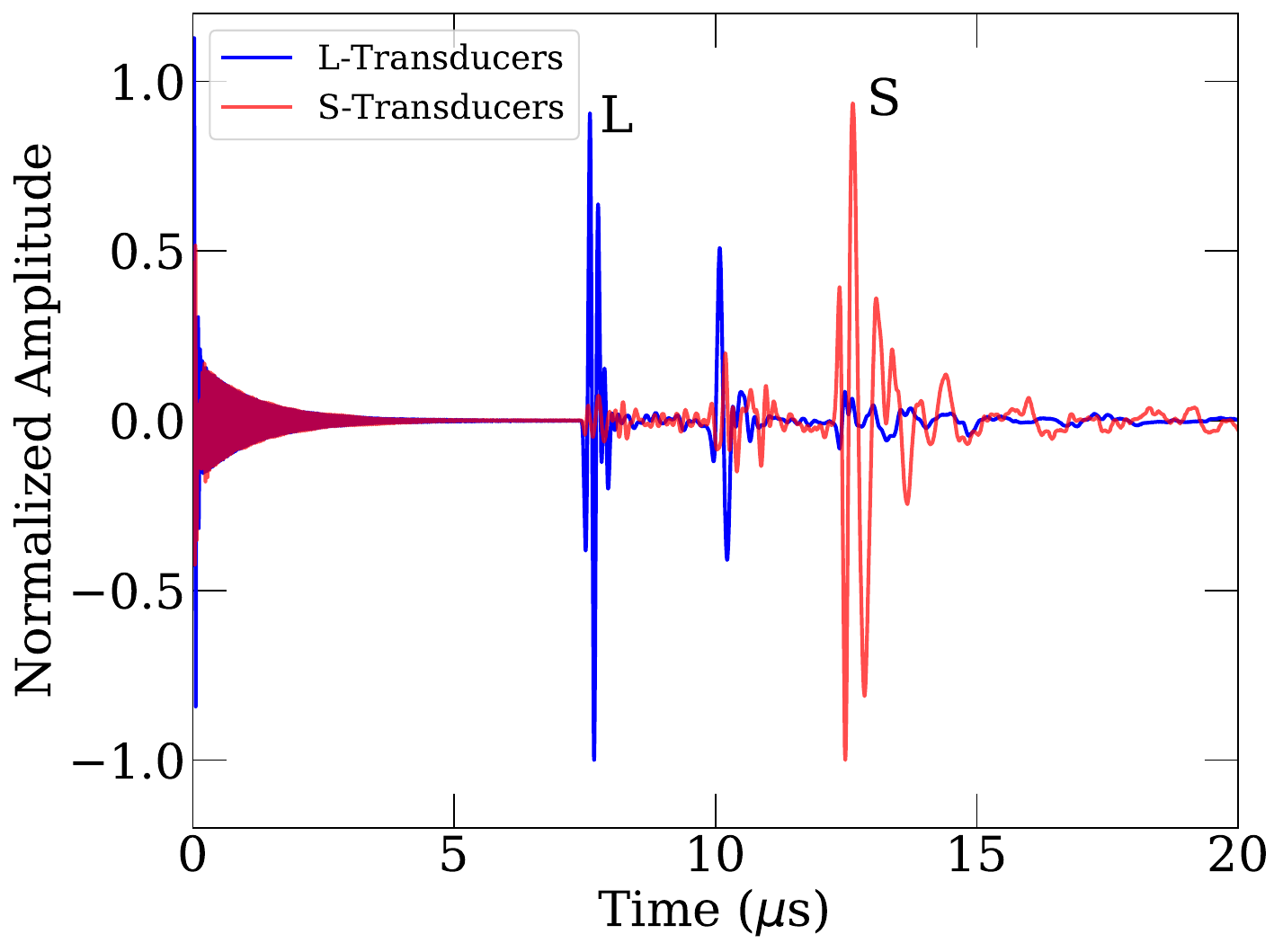} }     
\subfloat[Sample C]{ \includegraphics[width=0.48\linewidth]{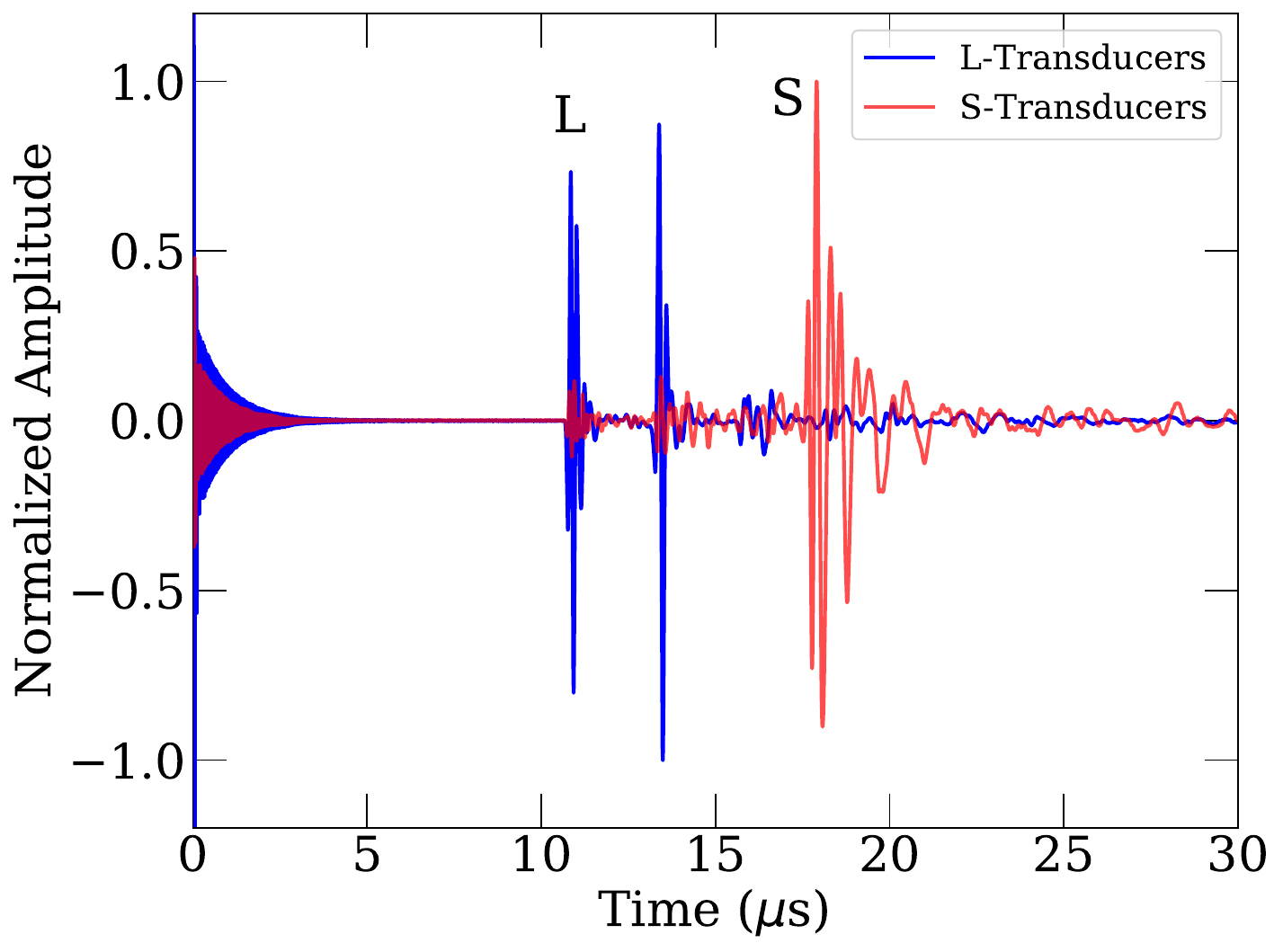} }
\caption{Comparison of waveforms produced by longitudinal and shear transducers through both samples at ambient conditions ($\sim 30\%$ RH), with the waveform of interest labeled (L) or (S) for longitudinal and shear, respectively. Blue: generated and recorded by longitudinal transducers, red: generated and recorded by shear transducers.}
\label{fig:Waveforms}
\end{figure}

\subsection{Ultrasonic Measurements}

The sample was held between two transducers, using 0.172-mm-thick nitrile rubber as a couplant, with a custom-built clamp assembly. The assembly contains three springs to press the transducers against the ends of the sample (see Appendix, Fig.~\ref{fig:Sample_Holder}). We used longitudinal and transverse contact transducer pairs (Olympus, part numbers V1091 and V157-RM, respectively) with fundamental frequency \SI{5}{MHz} and diameter \SI{6.35}{mm}. The source transducer was connected to a Pulser/Receiver (JSR Ultrasonics brand DPR300), which generates negative spike pulses 10 -- 70 ns in duration. The receiving transducer was used to record the transmitted signal. The pulse repetition rate was set to 100 pulses per second at 100 -- 250 V output. The pulser simultaneously sends a square wave trigger to the oscilloscope (Tekronix DPO2000B, 200 MHz, 1 GS/s, 1M record length) used to monitor the receiving transducer. Once the oscilloscope is triggered, defining $t=0$, the transmitted waveform was logged at a rate of 250 MHz. The oscilloscope was set to trigger at an amplitude of 1.4 V on the rising edge of the square trigger wave. The exact triggering amplitude had no effect on the time of flight. A single measurement consisted of averaging 512 transmitted waveforms to improve the signal to noise ratio. 

\subsection{Gravimetric Measurements}

While the ultrasonic measurements were performed on sample clamped between the transducers, the mass of the companion sample was concurrently monitored to determine the mass of water adsorbed. The companion sample sat on a watch glass adjacent to the mounted sample. After the relative humidity stabilized, and the time of flight was measured, the box was opened momentarily to remove the companion sample for weighing. The mass was measured on a Mettler AT261 DeltaRange analytical balance (weighing capacity 205 g), with a linearity of $\pm$ 0.15 mg. The time necessary to accurately measure the mass (seconds) was very short relative to the characteristic time for water vapor equilibration within the sample (hours). Initially, water uptake in sample C was monitored while measuring the sound speeds on sample V. After completing the adsorption and desorption branches, the samples were swapped and the experiments were repeated to obtain a complete data set.

\begin{figure}
\centering
\includegraphics[width=\linewidth]{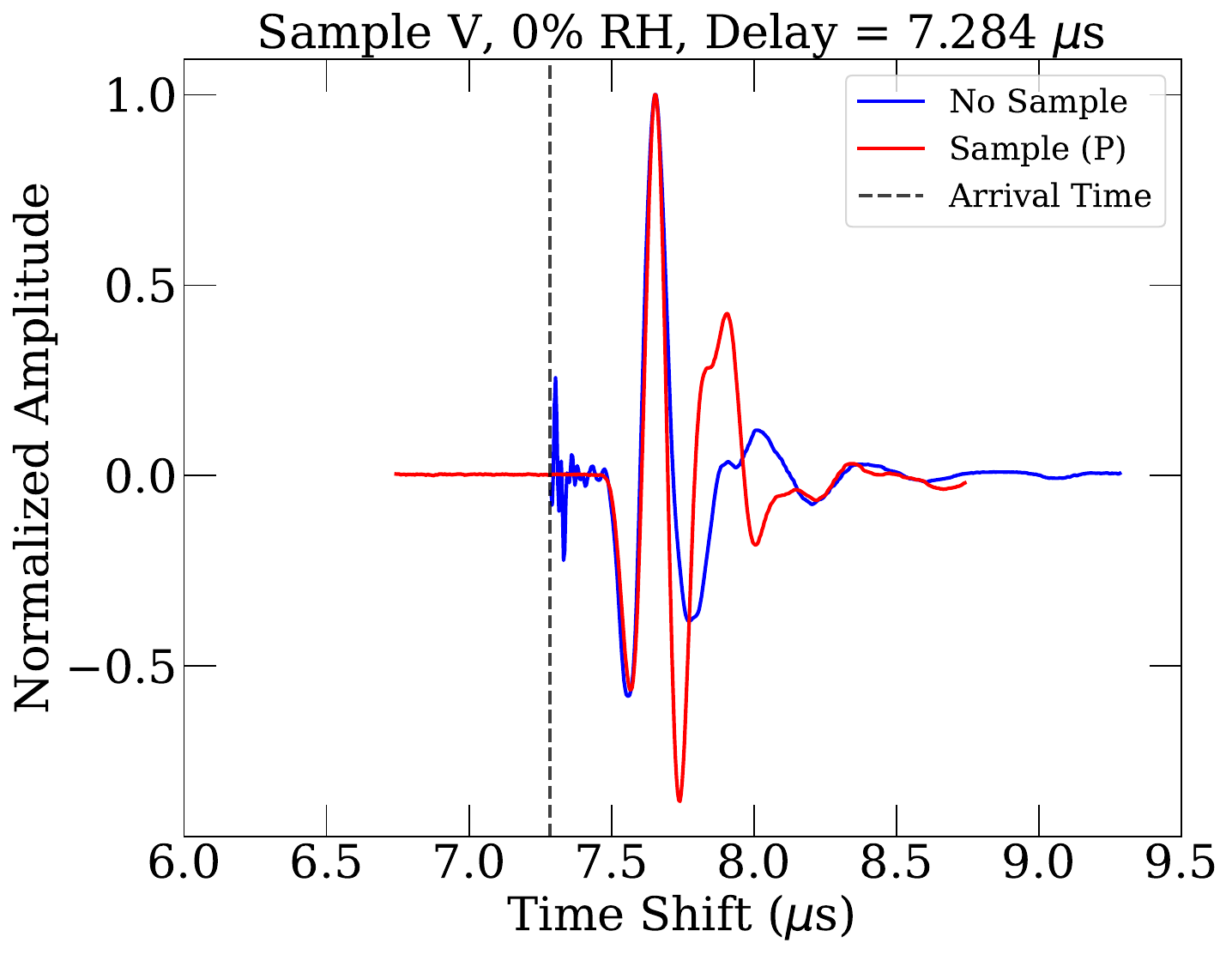}
\caption{Visual representation of the cross-correlation method, in which a reference waveform is shifted in the time domain to obtain the maximum similarity (correlation) between the two waveforms. The time shift that produces the maximum correlation coefficient represents the travel time used for calculating sound velocity. {The waveforms depicted in the figure are longitudinal waves produced by longitudinal transducers.}}
\label{fig:Cross_Correlation}
\end{figure}

\subsection{Data Analysis}

Typical waveforms recorded in this study are given in Fig.~\ref{fig:Waveforms}, which shows the signals produced by longitudinal and shear transducers. The time of flight is determined using a standard cross-correlation technique where the transmitted waves are correlated with a reference waveform. The latter is given by measuring the signal across two transducers clamped together, but separated by the nitrile couplant. Figure~\ref{fig:Cross_Correlation} shows a comparison of the waveform measured through a {porous} glass sample, and {the waveform} observed with the transducers clamped together after, {which was shifted} to maximize the correlation coefficient. The vertical dashed line indicates the time of flight determined from the cross correlation. The cross correlations were performed using the Python library ``scipy.signal.correlate", which follows a standard approach used in signal processing~\cite{Lyons2010}. The wave features between the longitudinal and shear waves seen in Fig.~\ref{fig:Waveforms} appear for samples which are smaller in diameter than the transducers, which is the case for the both samples considered here. This was additionally confirmed by performing the measurements on non-porous borosilicate glass samples of different diameters and lengths.

Another method to determine time of flight involved fitting the initial descent of the waveform to a linear function. The intersection of that line with the x-axis was defined as the arrival time. The two methods differed by $\sim \SI{10}{m \per s}$, corresponding to less then 0.3\%. For consistency, the results presented here rely on the cross correlation method.

The time of flight for the independent longitudinal and shear arrival are used to determine the corresponding sound speed given the sample length ($v_{\rm L}$ and $v_{\rm S}$, respectively). The sound speed is then combined with the sample density, $\rho$, to determine the longitudinal and shear moduli ($M$ and $G$, respectively):
\begin{equation}
    M = \rho {v_{\rm L}}^2, \qquad G = \rho {v_{\rm S}}^2.
\label{eq:Longitudinal_Modulus_Eq}
\end{equation}
These moduli are then used to calculate the bulk modulus, $K$ by the relation:
\begin{equation}
    K = M - \frac{4}{3}G.
\label{eq:Bulk_Modulus_Eq}
\end{equation}
This quantity represents the bulk modulus of the composite (nanoporous glass + adsorbed water). When the pores are completely filled with fluid, $K$ can be related to the bulk moduli of the constituents by the Gassmann equation~\cite{Gassmann1951,Gor2018Gassmann},
\begin{equation}
    K = K_0 + \frac{\left(1 - \frac{K_0}{K_{\rm s}}\right)^2}{\frac{\phi}{K_{\rm f}} + \frac{1-\phi}{K_{\rm s}} - \frac{K_0}{{K_{\rm s}}^2}},
\label{eq:Gassmann_Equation}
\end{equation}
where $\phi$ is the sample porosity, $K$ is the bulk modulus, and the subscripts $0$, ${\rm s}$, and ${\rm f}$ correspond to that of dry porous body, the solid phase, and the pore fluid, respectively. It is convenient to rearrange Eq.~\ref{eq:Gassmann_Equation} to solve for the quantity of interest, the bulk modulus of the confined fluid:
\begin{widetext}
\begin{equation}
    K_{\rm f} = \phi (K - K_0) \left[ \left(1-\frac{K_0}{K_{\rm s}}\right)^2 - \frac{(1-\phi)(K-K_0)}{K_{\rm s}} + \frac{K_0(K-K_0)}{{K_{\rm s}}^2} \right]^{-1}.
\label{eq:Rearranged_Gassmann}
\end{equation}
\end{widetext}

The parameter $K_{\rm s}$ is estimated using the modified Kuster and Toks{\"o}z (KT) effective medium theory~\cite{Kuster1974,Berryman1980,Sun2019}, which gives $K_{\rm s}$ and $G_{\rm s}$ when $K_0$, $G_0$, and $\phi$ are known. A summary of the KT theory used to determine $K_{\rm s}$ is in the Appendix. This theory is written for cylindrical pores, which is a reasonable approximation of the worm-like pores in the samples studied in this work.

{To measure the dry elastic properties, both samples were out-gassed at $\sim$\SI{E-5}{torr} for 24 hours before being transferred to a glove bag purged of ambient air and refilled with nitrogen. The ultrasound speed was measured using a combination of longitudinal and shear transducers, after which the samples were removed from the bag to measure their mass. Following these measurements, to drive off chemically adsorbed water from the pore surface, the samples were heated to \SI{200}{\degree C} at $\sim$\SI{E-5}{torr}. At these conditions, the samples were allowed to out-gas for 24 hours before cooling to room temperature. Ultrasonic and gravimetric measurements were then taken in the same manner as described above.}

{Following the aforementioned measurements, the samples were washed in a 30\% solution of hydrogen peroxide at \SI{80}{\degree C} to oxidize and remove trace organics from the pore surface. During washing, the samples lost their yellow tint and became clear. The samples were then transferred into deionized water to remove the peroxide solution. The above described vacuum measurements were repeated on the cleaned samples. We note here that the elastic properties of Vycor glass are sensitive to temperature. In particular, Scherer~\cite{Scherer1986} has shown that heat can irreversibly stiffen nanoporous glass. For this reason, we estimate the fluid bulk modulus using the dry mechanical properties obtained from the vacuum measurements described above, before the first heating cycle. The results of later measurements are provided in the Appendix.}

\section{Results}

\subsection{Nitrogen Gas Adsorption}

\begin{figure}[h]
\centering
\includegraphics[width=0.5\linewidth]{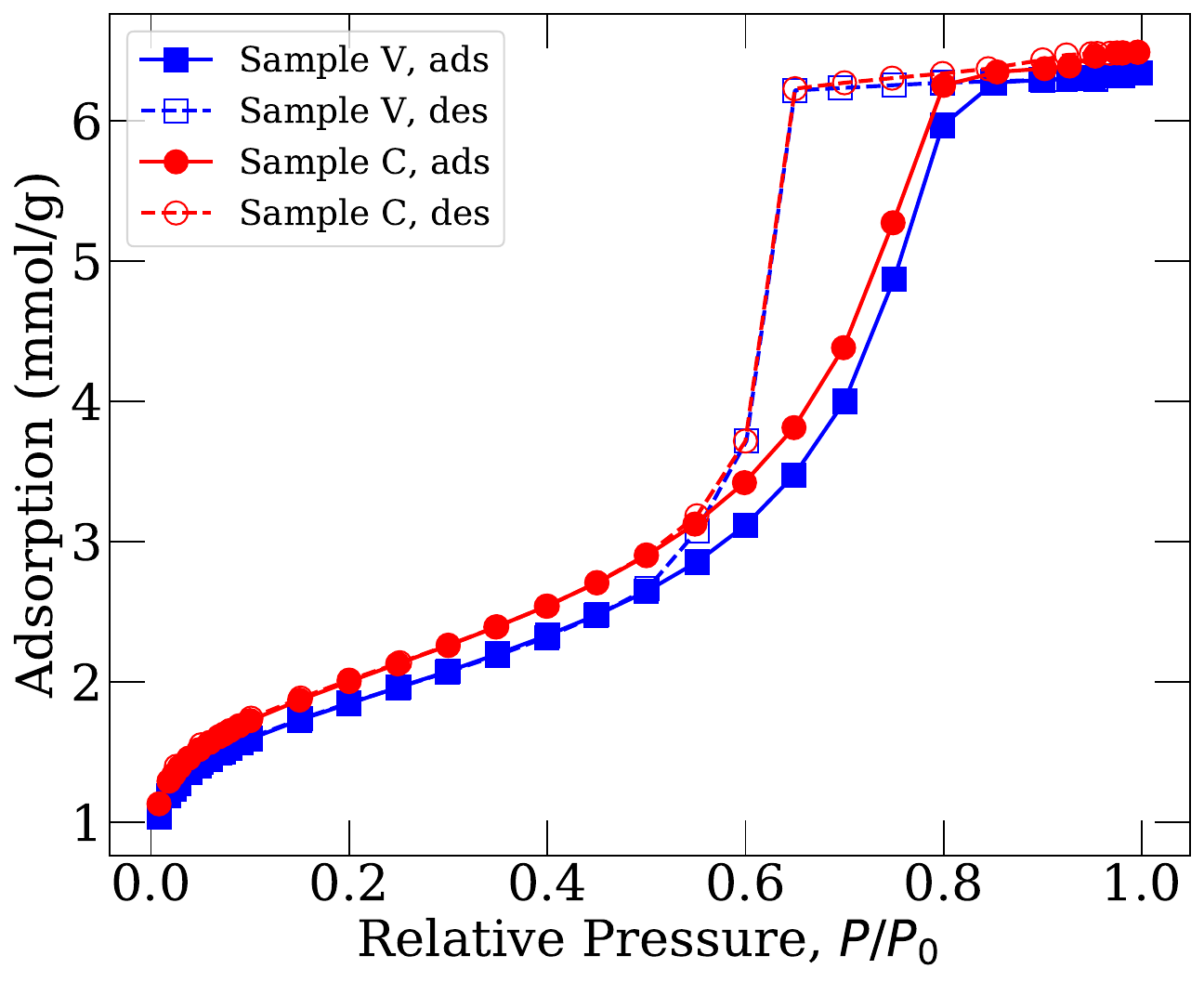}
\caption{Nitrogen adsorption isotherms on samples V and C measured at 77 K using a Autosorb-1C automated gas sorption system (Anton Paar). Filled markers represent adsorption, empty markers represent desorption.}
\label{fig:Nitrogen-Isotherms}
\end{figure}

\begin{figure}[h]
\centering
\includegraphics[width=0.6\linewidth]{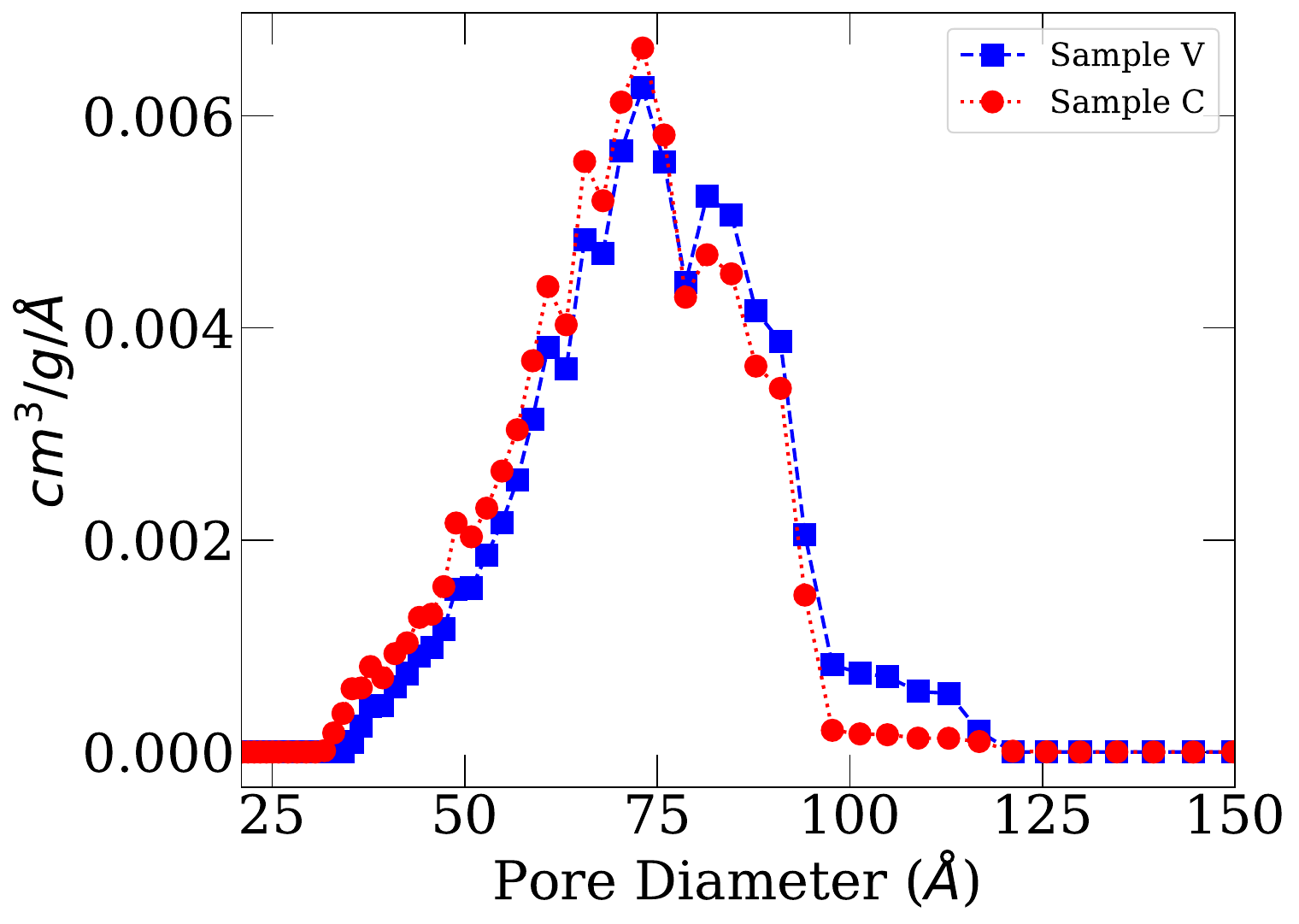}
\caption{Pore size distribution calculated from the nitrogen adsorption isotherms in Figure~\ref{fig:Nitrogen-Isotherms} for samples V and C. The differential distribution is a measure of the contribution of each pore size to the total pore volume per gram of each sample.}
\label{fig:Combined_PSD}
\end{figure}

The nitrogen adsorption isotherms are shown in Figure~\ref{fig:Nitrogen-Isotherms}, which indicates the responses are nearly indistinguishable {between the samples}. Prior to the capillary condensation, sample C adsorbs slightly more nitrogen than sample V, whereas the plateaus and the capillary evaporation regions are practically identical for the two samples. Abrupt, and nearly complete capillary evaporation occurs at the same relative pressure, $p/p_{0}\sim 0.6$, indicating that access to the porosity occurs through pores of a similar size.  Figure~\ref{fig:Combined_PSD} shows the pore size distribution (PSD) determined from the adsorption branch of the isotherm. Again, the pore size distributions of the samples are very similar, with the small difference observed near the tail corresponding to the largest pore diameters. 

\subsection{Water Adsorption Isotherms}

The relative mass of adsorbed water and saturation determined from the gravimetric measurements are shown in {Fig.~\ref{fig:Isotherms_Speed_Modulus}(A,B)}. Saturation is estimated as $s = \Delta m/\Delta m_{98}$, where $\Delta m$ is the change in sample mass at a given relative humidity, and $\Delta m_{98}$ is the change in sample mass at RH 98\%. Since the mass uptake is largely complete above RH=0.9, $s$ is a measure of the relative pore volume filled with water.

The isotherms are type IV, in agreement with the well-established adsorption data for water in Vycor glass~\cite{Amberg1952,Markova2000}. Vertical error bars represent the precision of the dry mass of both samples and of the balance used to measure the mass at subsequent relative pressures. The point at the origin is not measured, but it is reasonable to assume there is no adsorbed fluid at a relative pressure of zero. As humidity rises, the amount of water adsorbed gradually increases. In the region below the knee in the curve (e.g. $p/p_0 \leq$  0.2), a monolayer of water forms on the internal surface. Capillary condensation begins near a relative pressure of 0.75, where the water bridges the gap between the pore walls and fills the pores.

\subsection{Longitudinal and Shear Wave Speeds}

\begin{figure}[H]
\centering
\includegraphics[width=0.46\linewidth]{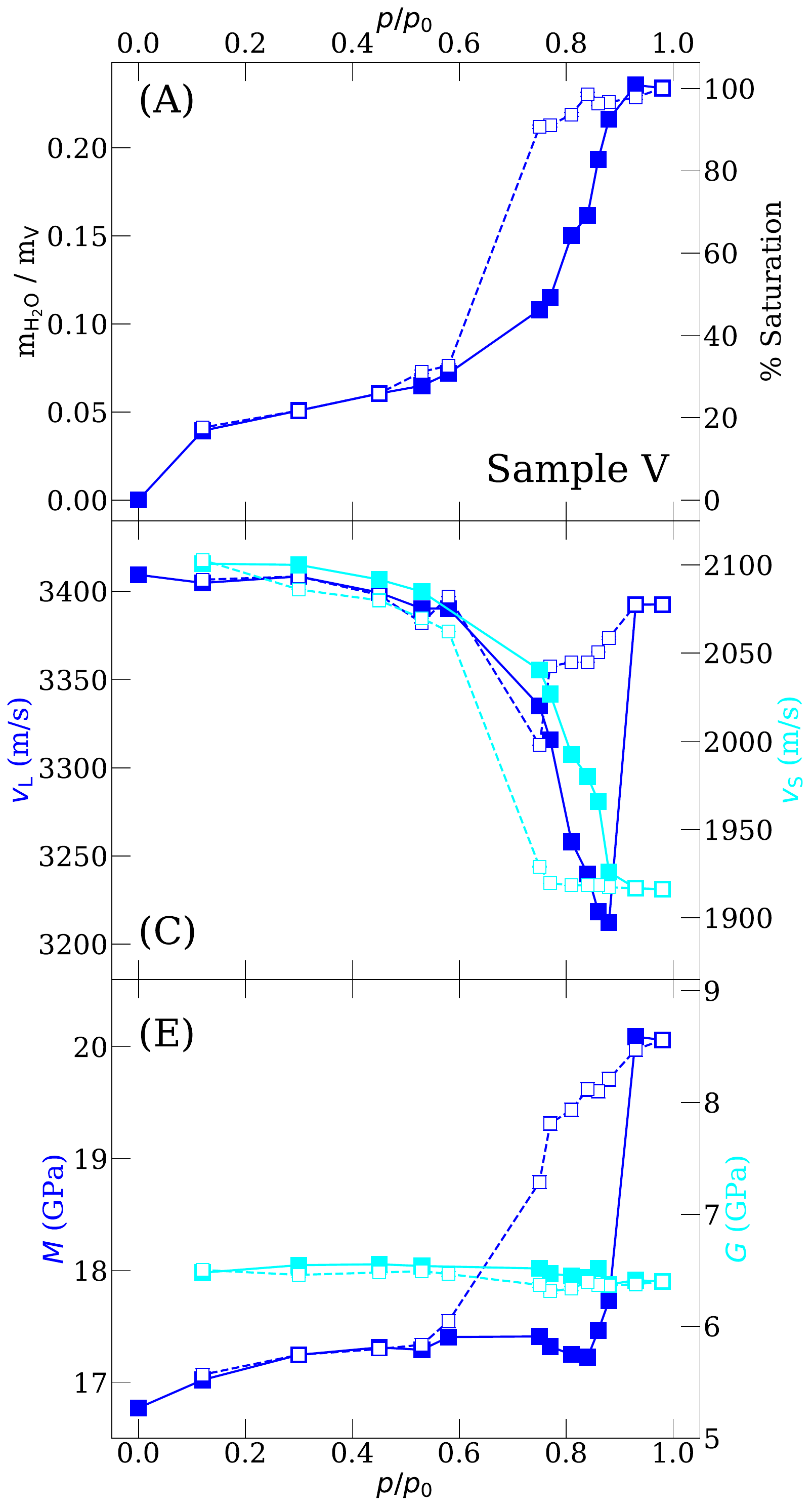}
\includegraphics[width=0.46\linewidth]{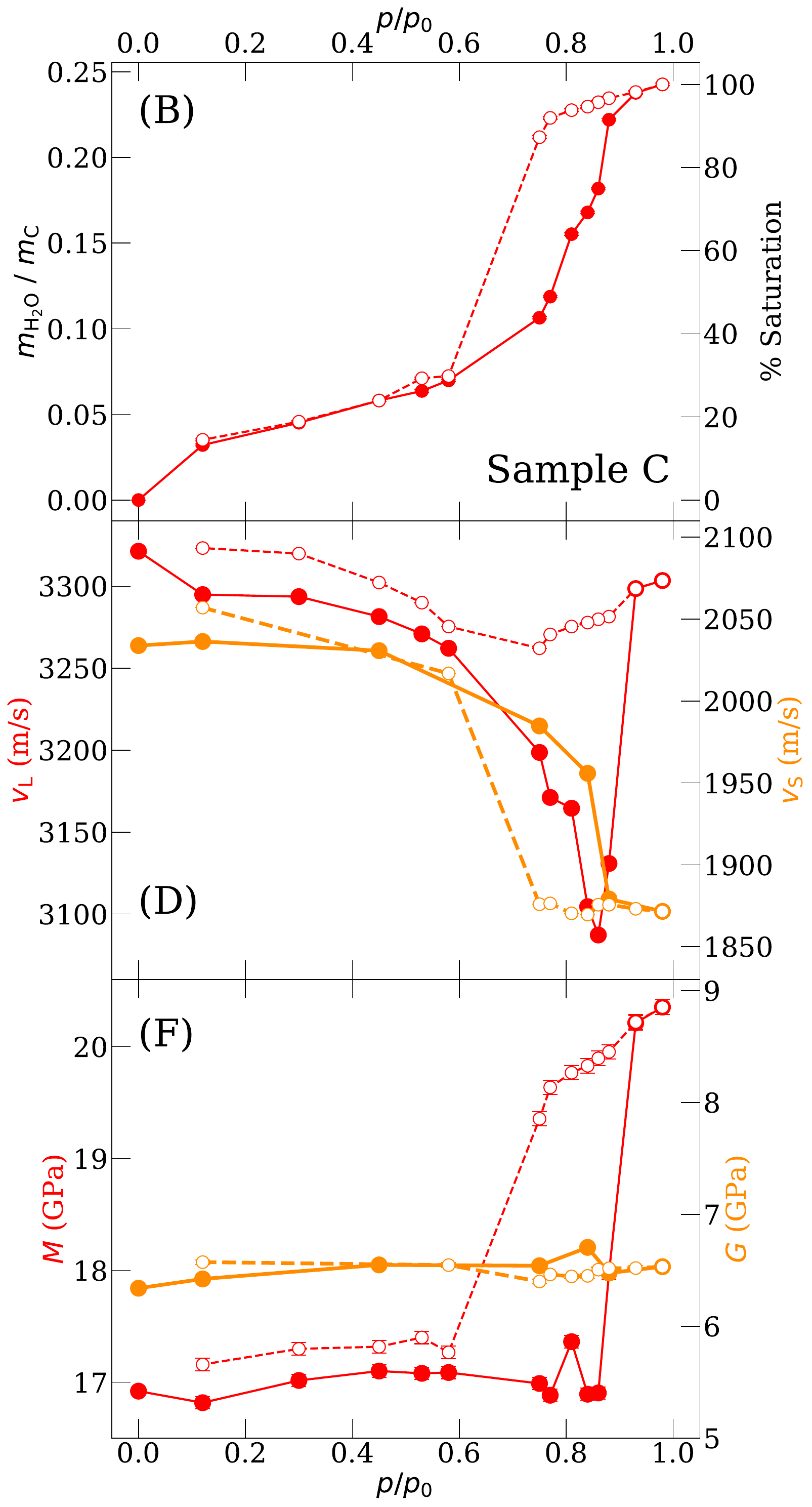}
\caption{{Results of ultrasonic and gravimetric measurements during adsorption (filled markers and solid lines) and desorption (empty markers and dashed lines). Lines connect adjacent data points as a guide to the eye. Sample V is represented in blue, sample C in red. Longitudinal speed and modulus share the same color as the adsorption isotherms. Shear speed and modulus use lighter color variants (cyan for sample V, orange for sample C). (A) and (B): Water adsorption isotherms, mass of water adsorbed, $m_{\rm H_2O}$ relative to the dry mass of each sample, $m_{\rm V}$ and $m_{\rm C}$. (C) and (D): Longitudinal, $v_{\rm L}$, and shear, $v_{\rm S}$, wave speeds. (E) and (F): Longitudinal and shear moduli calculated by Eq.~\ref{eq:Longitudinal_Modulus_Eq}. The speed and moduli presented at $p/p_0 = 0$ were obtained from the vacuum measurements explained in the methods section.}
}
\label{fig:Isotherms_Speed_Modulus}
\end{figure}

The longitudinal and shear wave speeds are shown in {Fig.~\ref{fig:Isotherms_Speed_Modulus}(C,D)} for both samples {(using the same vertical scale for both speeds)}. Each data point represents the average of up to 10 time of flight measurements taken several hours after the RH stabilized. As previously indicated, the speeds were determined from time of flight by cross-correlation with the waveform measured with no sample. The time of flight varied by no more than 0.1\% over several hours after the RH stabilized. The error bars account for this variation, and the error associated with measuring the sample length. In most cases the range of the error is smaller than the markers on the plot. 

Both samples exhibit similar behavior with respect to sound propagation throughout the adsorption and desorption processes. As RH increases, the sound speeds (both longitudinal and shear) gradually decrease, followed by a steeper decline with the onset of capillary condensation. The minimum in $v_{\rm L}$ is observed around $p/p_0 =$ 0.85, where the adsorbed fluid almost completely fills the pores. For longitudinal waves, this is followed by a marked increase in speed. In contrast, $v_{\rm S}$ levels off and continues gradually decreasing as maximum saturation is reached. At the beginning of desorption, $v_{\rm L}$ gradually decreases while $v_{\rm S}$ increases, following the same slope observed during the end of adsorption. As the hysteresis in mass adsorption closes, both speeds rise back toward the values observed at the lowest relative humidity. The increase in $v_{\rm S}$ is more dramatic than in $v_{\rm L}$, resembling the hysteresis loop observed in the adsorption isotherms.

{The longitudinal and shear wave speeds in sample C exhibit a hysteresis gap ($\sim$30 \si{m/s} and $\sim$20 \si{m/s}, respectively) that remains open below $p/p_0 =$ 0.6.} To ensure that this is not an artifact of the cross-correlation technique (e.g. caused by a difference in the shape of the waveforms), the sound speed was also determined by comparing only the initial descent of each waveform. With this method, the difference between adsorption and desorption decreased but remained noticeable ($\sim$20 \si{m/s}). {We note that prior to the experimental results presented in this work, sample V was used for initial adsorption and ultrasonic testing. During these tests, sample V was exposed to humidity swings between 12\% and 98\%. It is plausible that sample V had already undergone its adsorption-induced hysteresis.}

\subsection{Elastic Moduli}

\begin{figure}[h]
\centering
\includegraphics[width=0.48\linewidth]{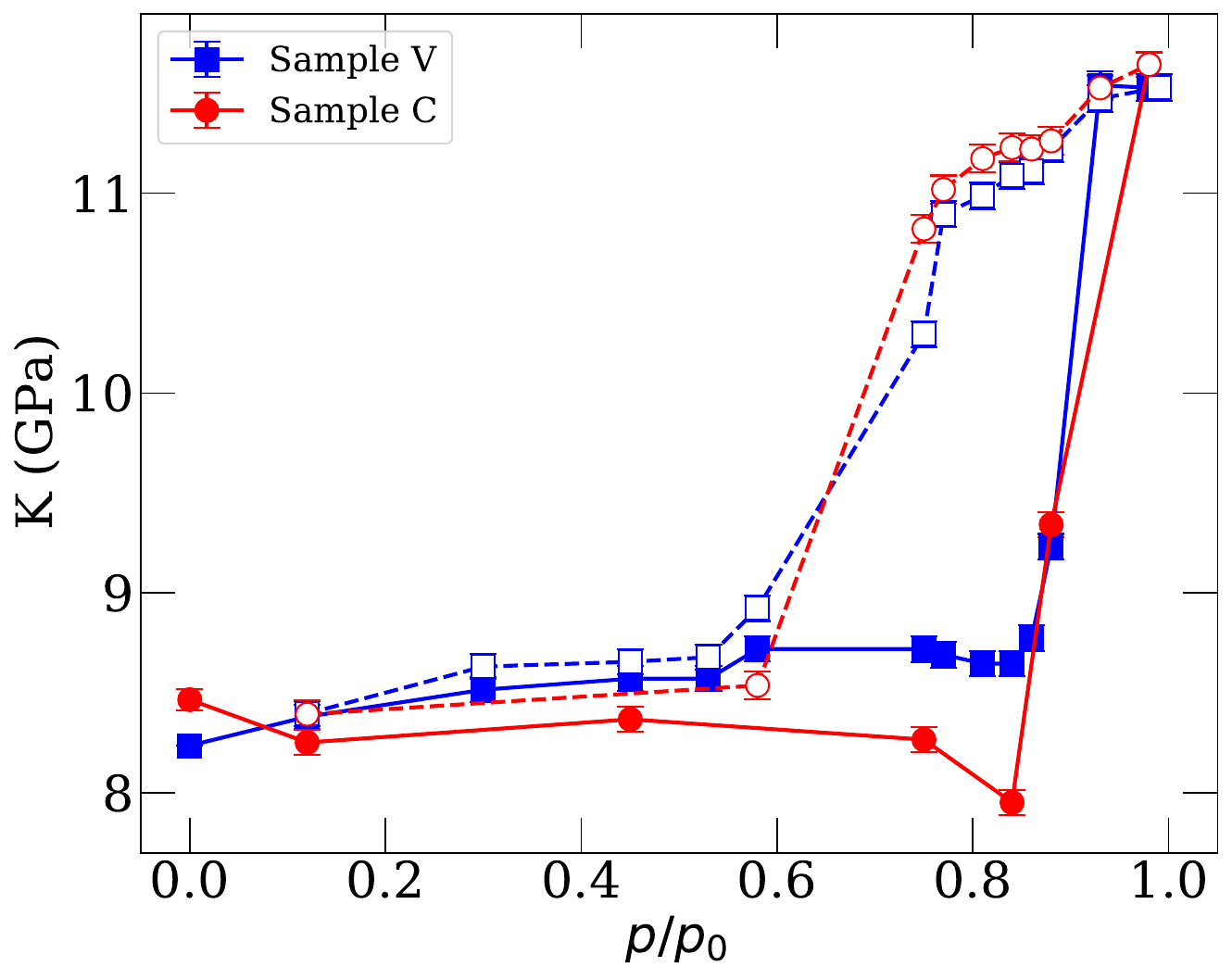}
\caption{Bulk modulus of samples V and C calculated by Eq.~\ref{eq:Bulk_Modulus_Eq}. Filled markers represent data on the adsorption branch, while empty markers represent desorption. The error bars are propagated from the uncertainty in the longitudinal and shear moduli.}
\label{fig:Bulk_Modulus}
\end{figure}

{Figs.~\ref{fig:Isotherms_Speed_Modulus}E and 6F} show the longitudinal and shear moduli of both samples {using the same vertical scale for both moduli}. {The moduli were} calculated using Eq.~\ref{eq:Longitudinal_Modulus_Eq}, where we account for changes in sample density associated with water vapor sorption. In contrast to observations concerning the sorption of hexane~\cite{Page1995} and argon~\cite{Schappert2018} in Vycor, the longitudinal modulus $M$ {of the glass sample with water} slightly increases at low relative pressures. The increase levels off near $p/p_0 \approx$ 0.6, followed by a decrease until $p/p_0 \approx$ 0.85. After capillary condensation, $M$ increases sharply, mimicking $v_{\rm L}$. On desorption, $M$ remains elevated until the sorption hysteresis loop closes, indicating evaporation of the pore fluid. This is the case for {longitudinal modulus of} both samples. In contrast, the shear modulus $G$ remains constant throughout the vapor sorption cycle. The bulk modulus $K$, calculated from Eq.~\ref{eq:Bulk_Modulus_Eq} and shown in Fig.~\ref{fig:Bulk_Modulus}, {exhibits similar behavior for both samples.}

\begin{figure}[h]
\centering
\includegraphics[width=0.7\linewidth]{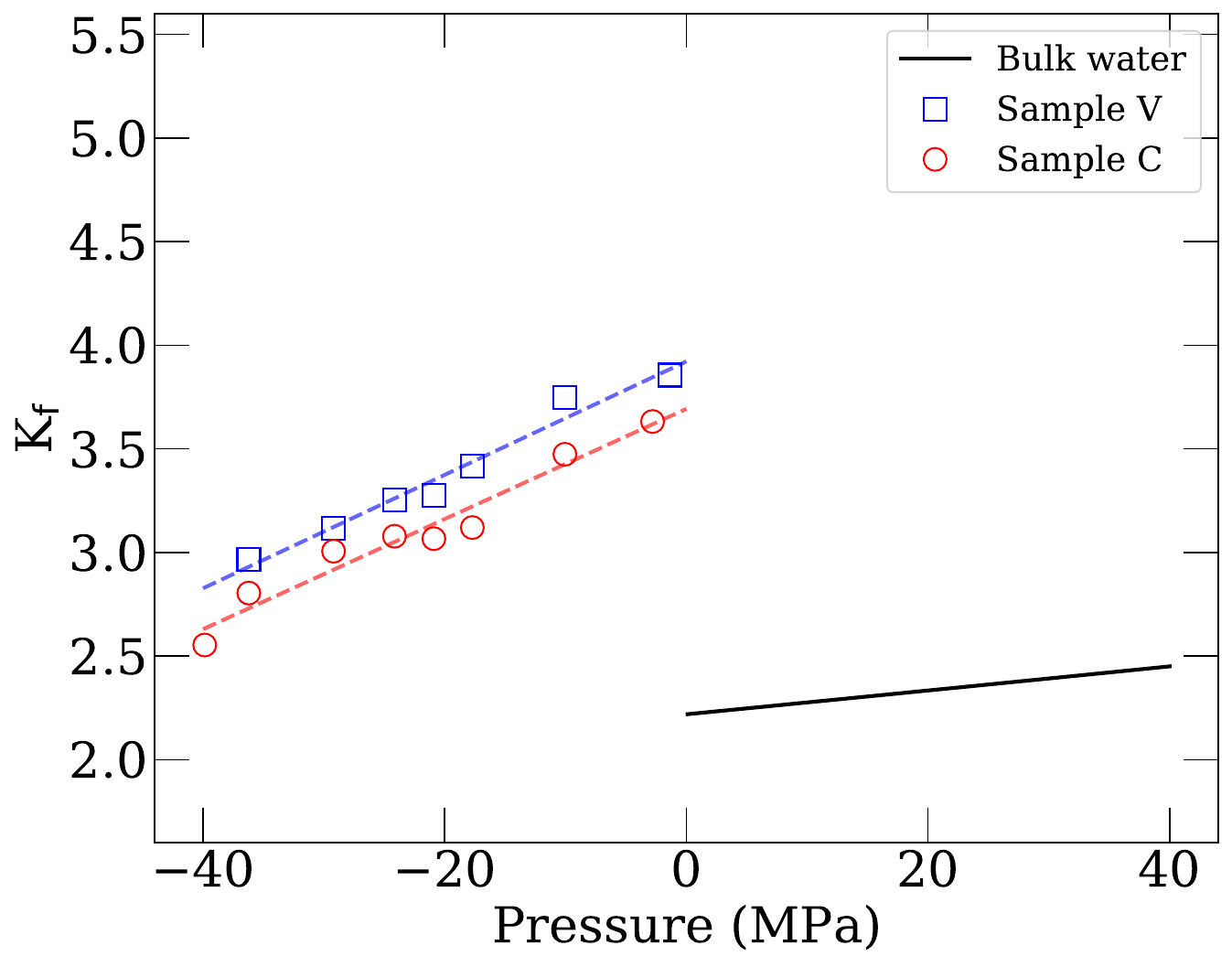}
\caption{Bulk modulus of water in Vycor glass calculated from the data in Fig.~\ref{fig:Bulk_Modulus} using Eq.~\ref{eq:Rearranged_Gassmann} plotted as a function of Laplace pressure (Eq.~\ref{eq:Laplace_Pressure}). The solid line represents the modulus of bulk water as a function of bulk pressure calculated based on Ref.~\cite{Wagner2002}. The dashed lines through the data markers are linear fits. {The negative part of the x-axis represents the Laplace pressure, $P_{\rm L}$, while the positive part of the x-axis represents bulk fluid pressure.}}
\label{fig:K_f_vs_RH}
\end{figure}

The values of $K$ observed while water fully occupies the pore volume (RH $> 0.85$ during the adsorption, and RH $> 0.75$ during desorption) are used to calculate the isothermal bulk modulus of confined liquid water using Eq.~\ref{eq:Rearranged_Gassmann}. The resulting fluid bulk modulus is shown in Fig.~\ref{fig:K_f_vs_RH}. We refer to this quantity as ``isothermal'' but note that the heat capacity ratio of liquid water at 1 \si{atm} and 300 \si{K} is 1.012~\cite{CoolPropSI}, making the difference between isothermal and adiabatic bulk modulus negligible. Previous studies%
~\cite{Schappert2014,Page1995,Dobrzanski2021} report a linear relation between a fluid's bulk modulus and the Laplace pressure, $P_{\rm L}$:
\begin{equation}
    P_{\rm L} = \frac{R_{\rm g}T}{V_{\rm m}} \ln{\left(\frac{p}{p_0}\right)},
\label{eq:Laplace_Pressure}
\end{equation}
where $R_{\rm g}$ is the ideal gas constant, $T$ is the absolute temperature, and $V_{\rm m}$ is the molar volume of the fluid. Figure~\ref{fig:K_f_vs_RH} shows the values of $K_{\rm f}$ determined in this work along with the isothermal bulk modulus of water in the bulk as a function of $P_{\rm L}$ or the bulk pressure, respectively. {$K_{\rm f}$ is calculated from Eq.~\ref{eq:Rearranged_Gassmann} using the values of $K_0$ and $K_{\rm s}$ listed in Table~\ref{tab:Mechanical_Props}}. The isothermal bulk modulus data for liquid water at 27 $^{\rm o}$C is calculated using the CoolProp Python library~\cite{CoolPropSI}, which utilizes the equation of state by Wagner and Pru\ss~\cite{Wagner2002}. It is worth noting that at relative pressures below 1.0, $P_{\rm L}$ is negative indicating the confined water is under suction. The data corresponding to positive pressure represents the isothermal bulk modulus of bulk water.

\begin{table*}
\caption{Calculated mechanical properties of samples V and C from ultrasound speed measurements {after out-gassing at \SI{E-5}{torr} for 24 hours without heating}. Other sets of values for Vycor 7930 are reported by Scherer et al. via sonic resonance and beam-bending, and Schappert and Pelster via a pulse-echo ultrasound method. $M_0$, $G_0$, $K_0$ and $\nu_0$ are the longitudinal, shear, and bulk moduli, and Poisson's ratio, respectively, of the dry porous material. $K_{\rm s}$ is calculated using the modified Kuster and Toks\"oz theory outlined in the Appendix.}
\label{tab:Mechanical_Props}
\begin{ruledtabular}
\begin{tabular}{lccccc}\\
& $M_0$ (\si{GPa}) & $G_0$ (\si{GPa}) & $K_0$ (\si{GPa}) & $\nu_0$ & $K_{\rm s}$ (\si{GPa}) \\
\colrule
Sample V     & 16.77 $\pm$ 0.06    & 6.40 $\pm$ 0.02    & 8.24 $\pm$ 0.06    & 0.192 $\pm$ 0.002 & 18.12 \\
Sample C     & 16.92 $\pm$ 0.06    & 6.34 $\pm$ 0.03    & 8.46 $\pm$ 0.06   & 0.200 $\pm$ 0.003 & 19.13 \\
Sonic Resonance~\cite{Scherer1986} & 15.64 \footnotemark[1] & 6.34 & 7.19 \footnotemark[1] & 0.15 & 15.37\\
Beam-bending~\cite{Vichit2000} & 15.9 & 6.08 & 7.8 & 0.16 & 14.3 \\
Pulse-Echo~\cite{Schappert2014} & 16.88 & 6.86 & 7.73 & 0.23 & 10.48 \\
\end{tabular}
\end{ruledtabular}
\footnotetext[1]{Scherer et al. report the Young's modulus, $E$, of their porous sample. Along with their reported shear modulus and Poisson's ratio, it is possible to calculate the longitudinal and bulk moduli.}
\end{table*}

\section{Discussion}

\subsection{Comparison of Water Adsorption Isotherms}

\begin{figure}[H]
\centering
\includegraphics[width=0.48\linewidth]{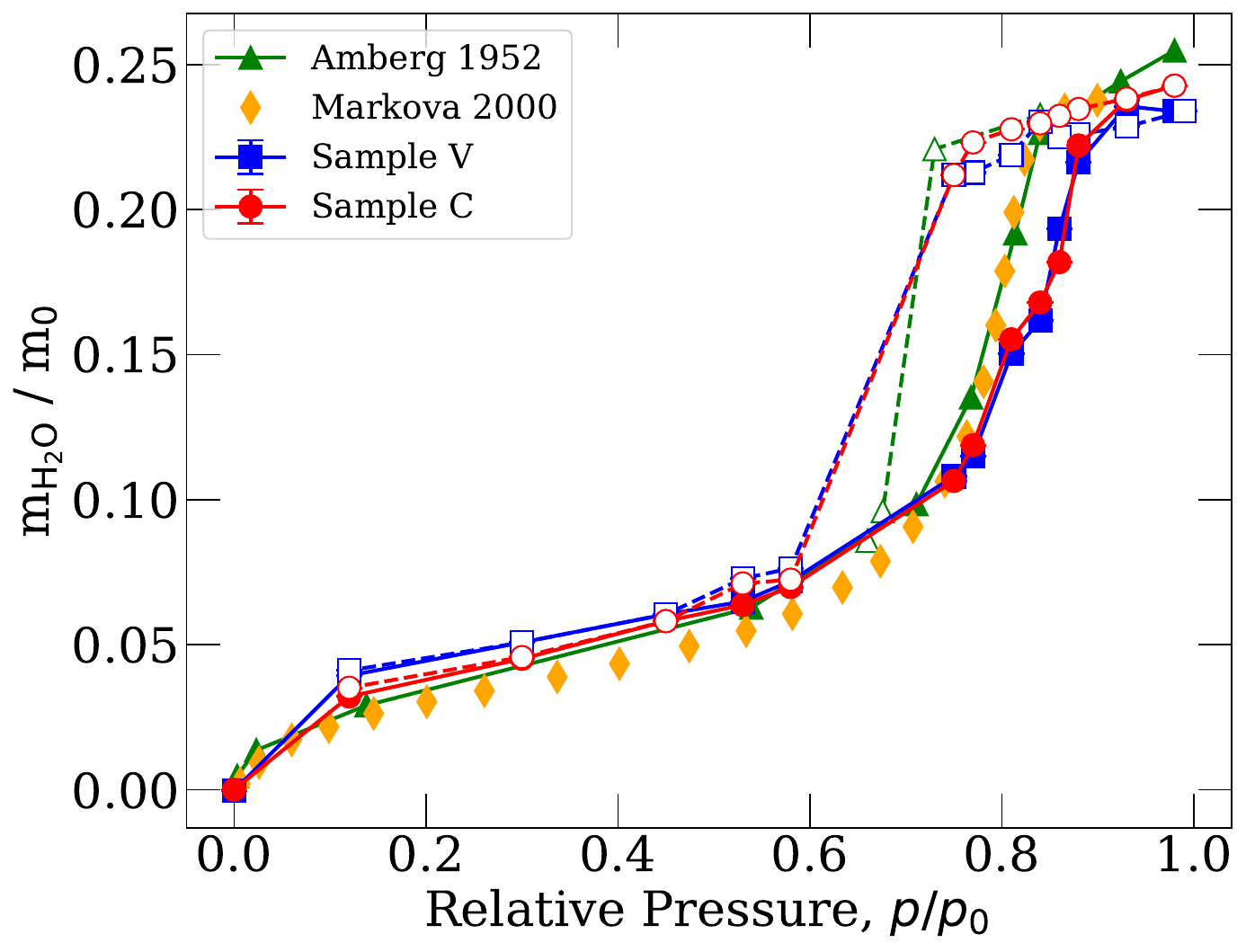}
\caption{Adsorption isotherms of water on nanoporous Vycor glass from two literature sources: by Amberg and McIntosh~\cite{Amberg1952} at \SI{25.8}{\degree C} and by Markova et al.~\cite{Markova2000} at \SI{25}{\degree C}, compared with the isotherms measured in this work at \SI{27}{\degree C}. Adsorption is represented by filled markers; desorption by empty markers. {The data from Markova et al. does not include desorption. Lines connecting markers are drawn to guide the eye.}}
\label{fig:Literature_Isotherms}
\end{figure}

{Figure~\ref{fig:Literature_Isotherms} includes data from previous works concerning water adsorption in Vycor. Overall, there is fairly good agreement along the adsorption branches. While Amberg and McIntosh observe a smaller hysteresis loop during desorption, we are unable to reproduce this in our samples due to the lack of an appropriate desiccant between 58 and 75\% RH. Furthermore, we observe good agreement between samples V and C, reflecting their similar composition and pore size distribution.}

\subsection{Applicability of the Gassmann Theory}

The Gassmann equation, which is used here to relate the measured modulus of the solid-fluid composite to the moduli of constituents, was originally derived for quasi-static deformations (e.g., the elastic properties must be probed by small perturbations). Therefore, we must consider the relevance of this theory to moduli inferred from the transmission of ultrasound. One constraint in the Gassmann theory is the average pore radius must be much smaller than the viscous skin depth of the fluid $\delta = \sqrt{\eta / \pi f \rho_{\rm f}}$, where $\eta$ and $\rho_{\rm f}$ are the dynamic viscosity and density of the pore fluid, respectively, and $f$ is the frequency of the propagating wave. The transducers used in this work have a nominal frequency of 5 MHz. We estimate the density of water in the pores at 100\% saturation as $\rho_{\rm f}\approx\SI{1044}{kg/m^3}$. Assuming the dynamic viscosity of liquid water, $\eta = \SI{8.509E-4}{\Pa\s}$ at 27 \si{\degree C} is unaffected by confinement, the viscous skin depth can be estimated as $\delta =\SI{228}{nm}$. For a more conservative estimate, we consider the maximum exhibited frequency of the transducers, 10 MHz, resulting in a viscous skin depth of $\delta = \SI{161}{nm}$. The pores in either sample are 7~nm, which is at least an order of magnitude smaller than the minimum viscous skin depth $\delta$.

Another stipulation of the Gassmann theory is that the constituent fluid has a shear modulus of zero, so that the shear modulus of the composite is not affected by fluid adsorption. In our case, the shear moduli of both samples are constant throughout sorption cycle. The Gassmann theory also requires  the frequency to be sufficiently low for the fluid pressure to equilibrate throughout the pore space within one wave period. This requires that the frequency be much lower than the characteristic frequency of the squirt-flow relaxation, which is on the order of $f_{\rm c} = \varrho^3 K_{\rm f}/\eta$, where $\varrho$ is the (smallest) pore aspect ratio in the system (see, e.g.,~\cite{Gurevich2022}, Chapter 6). For frequencies higher than $f_{\rm c}$, the shear modulus of the fully-saturated sample  should be significantly higher than for the dry sample~\cite{Mavko1991,Gurevich2022}. The fact that our measurements (as well as ultrasonic measurements on Vycor during adsorption of nitrogen~\cite{Warner1988} and argon~\cite{Schappert2014}) show a constant shear modulus in the entire range of liquid saturation from 0 to 100\% confirms that Vycor has no pores with very small aspect ratios (cracks). {Finally, we note that the Gassmann theory requires that the wavelength of the probing ultrasound be much smaller than the diameter of the porous sample. In our experiments, the wavelength never exceeds \SI{0.7}{mm}, which is more than nine times smaller than the diameter of either sample.}

\subsection{Comparison of Samples V and C}

The water adsorption isotherms in Fig.~\ref{fig:Isotherms_Speed_Modulus} are plotted as the mass of water adsorbed divided by the dry mass of each sample, so that the isotherms can be objectively compared. The samples exhibit close agreement until the end of capillary condensation on the adsorption branch ($p/p_0 \approx$ 0.85), after which the relative mass adsorbed in sample C is greater than that observed in sample V. On the desorption branch, this difference persists until the beginning of capillary evaporation ($p/p_0 \approx$ 0.75), where the hysteresis loop begins to close.

Qualitative agreement between the samples is also observed for the sound speeds, and the inferred moduli. Subtle quantitative differences can be attributed to the initial properties of both samples. For instance, the longitudinal and shear moduli are similar but $v_{\rm L}$ and $v_{\rm S}$ differ between each sample by more than 100 \si{m/s} due to the samples having different solid densities (see Table~\ref{tab:Samples}).

\subsection{Mechanical Properties of Dry Samples and Confined Water}

In Table~\ref{tab:Mechanical_Props}, the calculated mechanical properties of samples V and C are presented alongside those reported by Scherer et al. for the same material via sonic resonance~\cite{Scherer1986} and by a beam-bending method~\cite{Vichit2000}, as well as Schappert and Pelster who used pulse-echo ultrasound measurements~\cite{Schappert2014}. While the shear modulus of both samples agrees with the literature sources, we observed longitudinal moduli higher than any of the listed values. This leads to higher values for $K_0$ and $K_{\rm s}$.
Table~\ref{tab:Mechanical_Props} also shows the Poisson's ratio $\nu_0$ of the dry samples:
\begin{equation}
    \nu_0 = \frac{M_0 - 2G_0}{2M_0 - 2G_0}.
    \label{eq:nu}
\end{equation}

Note that the bulk modulus of the fluid from Eq.~\ref{eq:Rearranged_Gassmann} is sensitive to the values of $K_0$ and $K_{\rm s}$, where the latter is the main source of potential uncertainties. If the dry elastic constants reported by Scherer et al. and Schappert and Pelster are used to calculate $K_{\rm s}$ of Vycor glass via the modified KT theory, $K_{\rm f}$ inferred from the Gassmann theory will differ from the results in Figure~\ref{fig:K_f_vs_RH}, as discussed in Ref.~\cite{Gor2018Gassmann}. {We find that a hypothetical underestimate of $K_{\rm s}$ by 10\% or 20\% causes an overestimate of $K_{\rm f}$ by 20\% and 50\%, respectively.}

Irrespective of the values for the $K_0$ and $K_{\rm s}$, the fluid bulk modulus is linearly dependent on the pressure of the fluid in the pores (represented here by the Laplace pressure for simplicity)~\cite{Gor2010,Gor2014}, which is consistent with the Tait-Murnaghan equation~\cite{Murnaghan1944,Birch1952}:
\begin{equation}
    K_{\rm f}(P_{\rm L}) \approx K_{\rm f}(0) + \alpha P_{\rm L},
\label{eq:Tait-Murnaghan}
\end{equation}
where $\alpha = {\rm d}K_{\rm f}/{\rm d}P_{\rm L}$. Previous molecular simulation studies suggested that $\alpha$ for argon or nitrogen in silica pores is similar to that for bulk liquid argon~\cite{Gor2015compr, Gor2016Tait, Dobrzanski2018}, or liquid nitrogen \cite{Maximov2018}, respectively. However, Figure~\ref{fig:K_f_vs_RH} indicates confined water exhibits a different $\alpha$ than that exhibited by bulk water ($\alpha = 5.76$). In particular, $\alpha \sim 25.6$ for sample V and $\alpha \sim 27.8$ for sample C. These deviations are too high to be explained by uncertainties in the modulus $K_{\rm s}$ used for calculation of $K_{\rm f}$ from the experimental data. 

Molecular simulation parametric studies of confined argon showed that when the solid-fluid interactions are stronger, the parameter $\alpha$ increases~\cite{Gor2016Tait}. It could be that confinement affects the parameter $\alpha$ for water differently than for simple fluids, such as nitrogen or argon. More insight into $\alpha$ of confined water can be obtained from experiments and simulations of water in nanopores of various compositions.

\section{Conclusion}

To investigate the compressibility of confined water, we performed water vapor sorption experiments on two nanoporous glass samples and concomitantly monitored the effect on longitudinal and shear sound speed. Using the measured density of the water laden glass, we infer the longitudinal and shear moduli as a function of relative humidity. The longitudinal modulus varied with relative humidity, while the shear modulus did not change. Therefore, we utilized the Gassmann theory to infer the bulk modulus of water confined in the saturated glass. For both of the porous samples, we find that the modulus of confined water is consistently greater than that of bulk liquid water at the same temperature. Furthermore, the modulus of the confined water exhibited a linear dependence on Laplace pressure. Altogether these observations are consistent with previously reported data on non-polar fluids, such as nitrogen or argon, suggesting that confinement causes significant stiffening for any fluid. Note, modulus increase reported here is more pronounced than that reported for the noted non-polar fluids confined to the same nanoporous glass.

This work should assist with the progression towards borehole based ultrasonic measurements to probe natural nanoporous media, such as shale and coal, in order to detect specific confined fluids. While we investigated the effects of confinement on one fluid found in subsurface reservoirs, additional work is necessary to yield a useful measurement technique. In particular, similar studies using other relevant fluids, like hydrocarbons or salt solutions, or nanoporous media with a range of physical properties would aid in assessing the utility of ultrasound to excite a characteristic response. 

\section{Data Accessibility}

{The raw data for Figures 4-9 and A2 are available as a GitHub repository~\cite{Raw_data}. Each file is written in plain text (.txt) or comma-separated values (.csv), and named for each respective figure.}

\begin{acknowledgments}
G.Y.G., A.K. and A.F.K. thank the support from New Jersey Institute of Technology Seed Grant, and NSF CBET-2128679 grant. J.O. thanks the NJIT Provost URI Summer Research Fellowship and the McNair Postbaccalaureate Achievement Program. {We thank George Scherer for providing a number of comments that helped improve the manuscript.}

\end{acknowledgments}

\appendix

\renewcommand{\thefigure}{A\arabic{figure}}
\setcounter{figure}{0}

\renewcommand{\thetable}{A\arabic{table}}
\setcounter{table}{0}

\renewcommand{\theequation}{A\arabic{equation}}
\setcounter{equation}{0}

\section{Additional Experimental Details}

\subsection{{Estimates of Dry Properties After Vacuum Out-Gassing}}

{The following tables show the results of ultrasound speed measurements to calculate the elastic properties of samples V and C after out-gassing at a pressure of \SI{E-5}{torr}, as outlined in the methods section of the main text. The samples were first out-gassed at room temperature, then at \SI{200}{\degree C}. Since darkening of the samples was observed after heating, the samples were then washed, and the above measurements were repeated. The tables also show the resultant bulk modulus of water from Eq.~\ref{eq:Rearranged_Gassmann}, using the corresponding dry and solid bulk moduli. The values obtained from the ``pre-wash, ambient'' (not heated, not washed) measurements are used for calculating $K_{\rm f}$.}

{After the experiments presented in the main text, one of the shear ultrasonic transducers became inoperable. For this reason, the travel time of shear waves needed to be recorded using the pulse echo method. In sample V, this proved difficult due to its shorter length and the presence of noise in the signal, but the travel time of both longitudinal and shear waves could be measured successfully in sample C. Further, during the second ambient vacuum cycle, sample V developed a fracture along the axial direction near the center of the sample. Attempts at ultrasonic measurements on this sample were unsuccessful. To prevent additional damage, sample V was not heated for a second time. Throughout these experiments, sample C remained intact and permissive to ultrasonic testing.}

{For the reasons explained above, in Table~\ref{tab:Vacuum_Table_V}, the shear modulus of sample V for the pre- and post-wash measurements is estimated based on the relative change in the shear modulus of sample C. The same estimation is made for the longitudinal modulus of the post-wash measurements. Both samples exhibited quantitatively similar relative changes of these properties during water adsorption.}

\begin{table}[H]
\caption{{Calculated mechanical properties of sample V from ultrasound speed measurements after following the procedure outlined in the main text. $K_{\rm s}$ is calculated using the modified Kuster and Toks\"oz theory outlined in the Appendix. $K_{\rm f}$ is calculated using Eq.~\ref{eq:Rearranged_Gassmann}, with the respective dry and solid moduli.}}
\label{tab:Vacuum_Table_V}
\begin{ruledtabular}
\begin{tabular}{lcccccc}\\
& $M_0$ (\si{GPa}) & $G_0$ (\si{GPa}) & $K_0$ (\si{GPa}) & $\nu_0$ & $K_{\rm s}$ (\si{GPa}) & $K_{\rm f}$ (\si{GPa}) \\
\colrule
12\% RH & 17.02 & 6.48 & 8.38 & 0.193 & 18.46 & 3.635 \\
Pre-wash, ambient & 16.77 & 6.40 & 8.24 & 0.192 & 18.12 & 3.855 \\
Pre-wash, \SI{200}{\degree C} & 17.54 & 6.95 & 8.27 & 0.172 & 17.42 & 4.114 \\
Post-wash, ambient & 16.45 & 6.40 & 7.91 & 0.181 & 16.98 & 4.561 \\
Post-wash, \SI{200}{\degree C} & 17.72 & 6.90 & 8.51 & 0.181 & 18.26 & 3.609 \\
\end{tabular}
\end{ruledtabular}
\end{table}

\begin{table}[H]
\caption{{Calculated mechanical properties of sample C from ultrasound speed measurements after following the procedure outlined in the main text. $K_{\rm s}$ is calculated using the modified Kuster and Toks\"oz theory outlined in the Appendix. $K_{\rm f}$ is calculated using Eq.~\ref{eq:Rearranged_Gassmann}, with the respective dry and solid moduli.}}
\label{tab:Vacuum_Table_C}
\begin{ruledtabular}
\begin{tabular}{lcccccc}\\
& $M_0$ (\si{GPa}) & $G_0$ (\si{GPa}) & $K_0$ (\si{GPa}) & $\nu_0$ & $K_{\rm s}$ (\si{GPa}) & $K_{\rm f}$ (\si{GPa})\\
\colrule
12\% RH & 16.82 & 6.42 & 8.25 & 0.191 & 18.24 & 4.074 \\
Pre-wash, ambient & 16.92 & 6.34 & 8.46 & 0.200 & 19.13 & 3.621 \\
Pre-wash, \SI{200}{\degree C} & 18.05 & 6.89 & 8.87 & 0.192 & 19.79 & 3.127 \\
Post-wash, ambient & 16.93 & 6.35 & 8.47 & 0.200 & 19.13 & 3.616 \\
Post-wash, \SI{200}{\degree C} & 18.24 & 6.84 & 9.11 & 0.200 & 19.13 & 3.116 \\
\end{tabular}
\end{ruledtabular}
\end{table}

\subsection{Ultrasonic Attenuation}

\begin{figure}[H]
\centering
\includegraphics[width=0.48\linewidth]{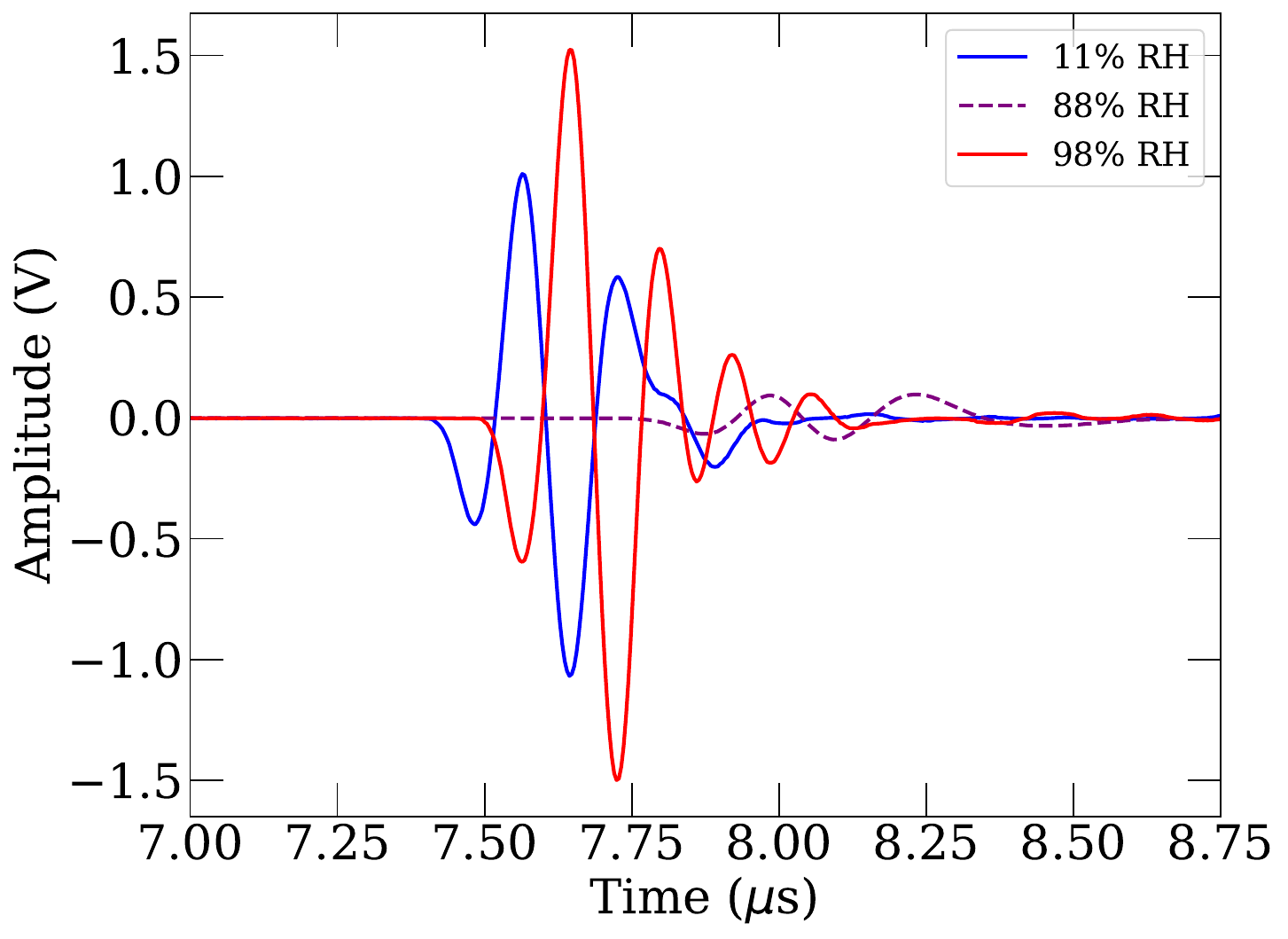}
\caption{{Longitudinal waves produced by longitudinal transducers} recorded through sample V at three humidities during adsorption, illustrating the change in travel time and the observed attenuation. The wave which exhibits maximum attenuation is drawn in dashed lines.}
\label{fig:Attenuation_Waves}
\end{figure}

In certain regions of $p/p_0$, the amplitude of measured ultrasound waves dropped precipitously, followed by recovery. In both samples, this occurred at 0.84 -- 0.88 on adsorption and 0.88 -- 0.75 on desorption, with the former being more prominent than the latter. During desorption, attenuation began at $p/p_0 = 0.88$, gradually increased and peaked at 0.75, just before the pores emptied. Figure \ref{fig:Attenuation_Waves} exemplifies the waveforms at three distinct values of humidity. The attenuation was observed qualitatively for both longitudinal and shear waves. On desorption, at the same time as attenuation was observed, the samples became white and opaque to light.

\begin{figure}[h]
    \centering
    \subfloat[Attenuation]{ \includegraphics[width=0.48\linewidth]{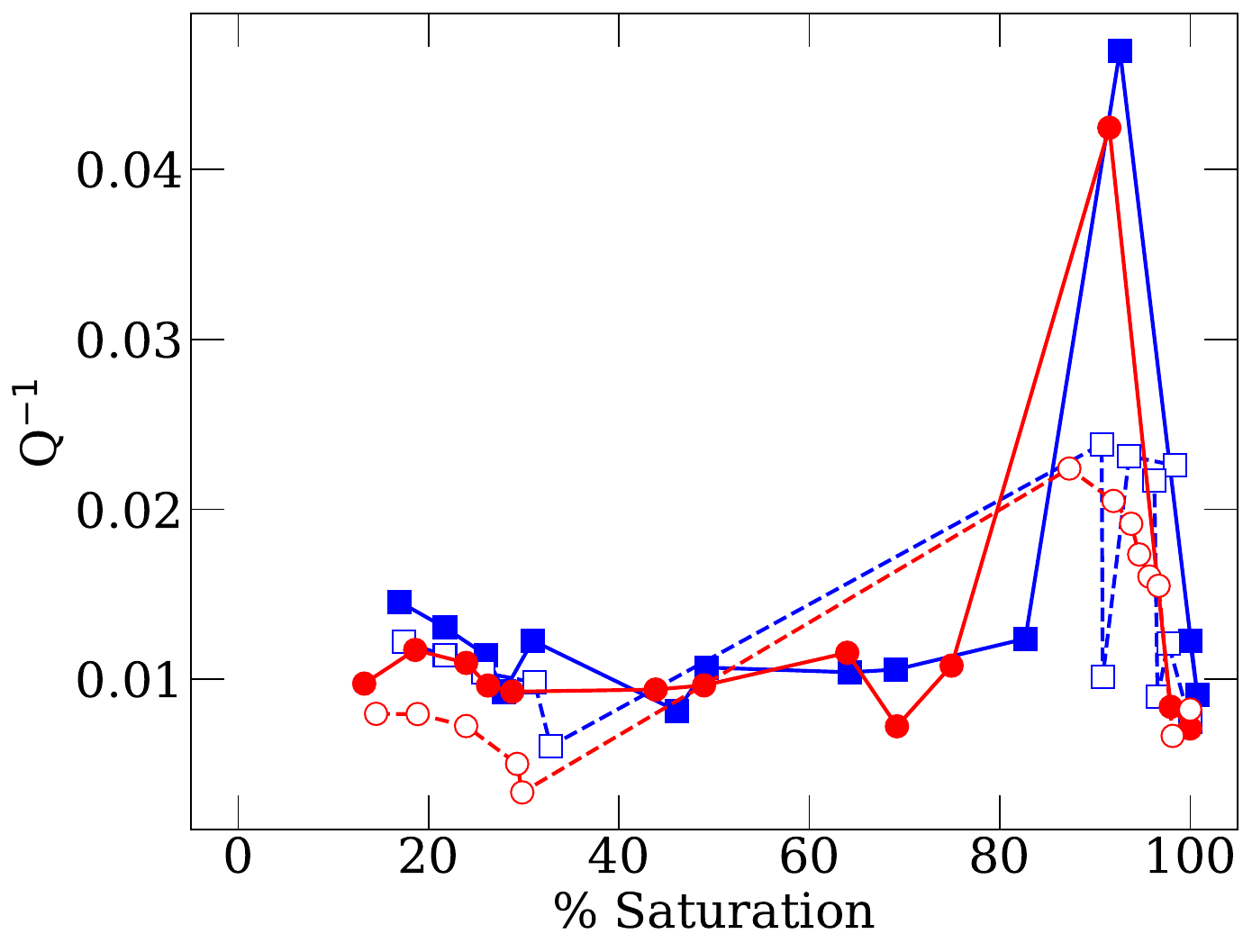} }
    \subfloat[Bulk modulus]{ \includegraphics[width=0.48\linewidth]{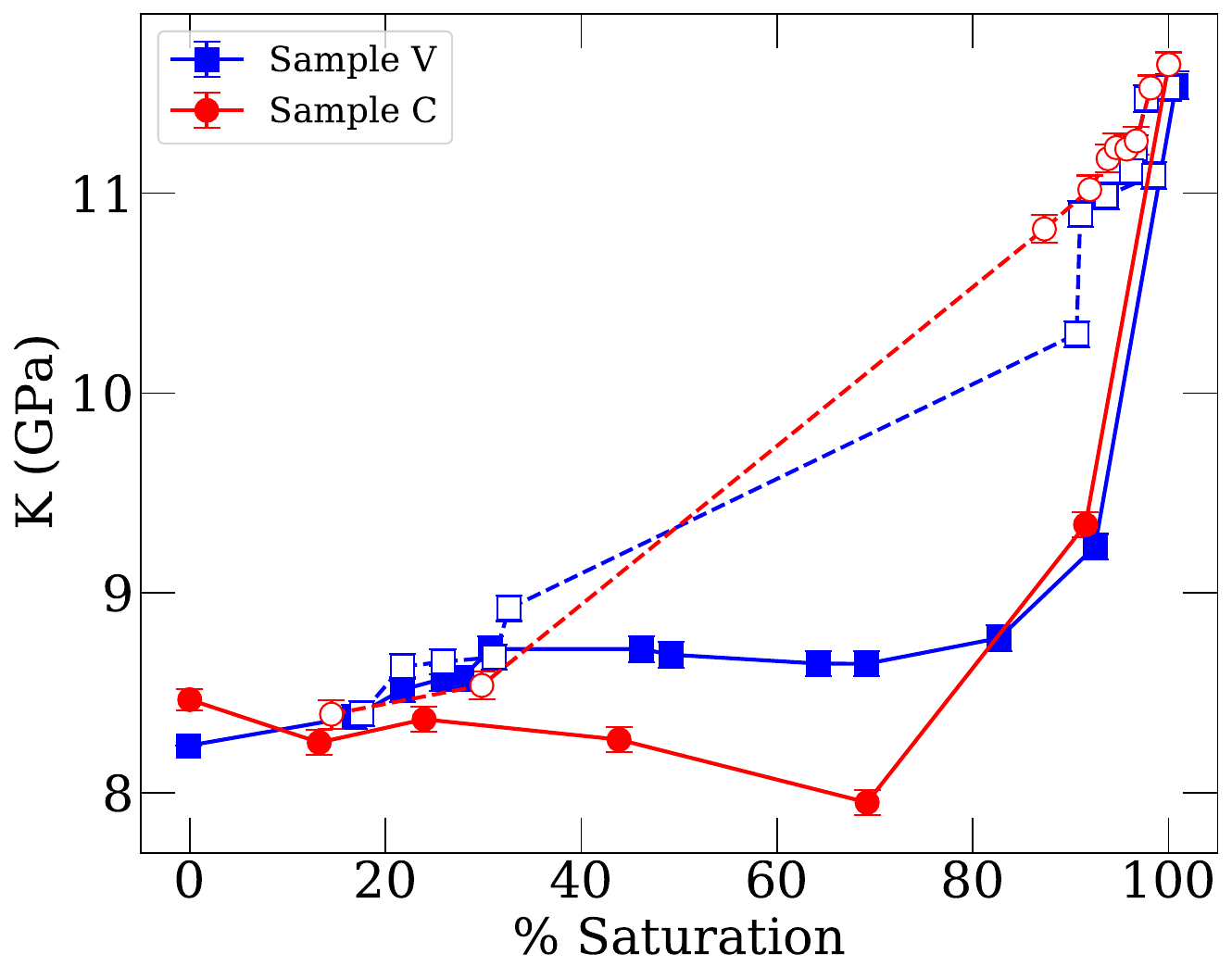} }
    \caption{(a) Attenuation of longitudinal waves through samples V (blue) and C (red) at different levels of saturation. (b) Bulk modulus of samples V and C calculated by Eq.~\ref{eq:Bulk_Modulus_Eq} as a function of the sample saturation. Filled markers represent data on the adsorption branch, while empty markers represent desorption. The error bars are propagated from the uncertainty in the longitudinal and shear moduli.}
    \label{fig:Attenuation}
\end{figure}

During capillary condensation ($p/p_0 = 0.88$, adsorption) and emptying ($p/p_0 = 0.75$, desorption), the amplitude of ultrasound waves shows a significant decrease in both samples. Fig.~\ref{fig:Attenuation} (a) shows the dimensionless attenuation ($Q^{-1}$) computed using the spectral ratio method~\cite{Sears1981} as a function of water saturation. {$Q^{-1}$ is a dimensionless measure of attenuation commonly used in the theory of viscoelasticity and geophysics, and corresponds to the energy loss per one wave period (as opposed to dimensional measures such as the attenuation coefficient, which represents energy loss per meter)~\cite{Knopoff1964, Winkler1982, McCann2009}.} On adsorption, $Q^{-1}$ shows a sharp peak around water saturation of 90\%. This behavior of attenuation is similar to that during hexane adsorption in Vycor glass~\cite{Page1995}. While water and hexane are significantly different fluids, the results of this work are qualitatively similar to those observed by Page et al., and it is reasonable to consider that the causes are the same as well. Moreover, this behavior is very similar to the behavior of elastic-wave attenuation in macro-porous media partially saturated with water~\cite{Cadoret1998} at relatively low frequencies. This attenuation is usually attributed to the relaxation of fluid pressure between portions of the sample with different levels of liquid saturation~\cite{Kobayashi2016} and indicates that, even on adsorption, the saturation within the sample shows slight spatial variations. These variations cause spatial variations in compressibility, and thus also variations in fluid pressure in the passing wave. Fluid pressure gradients, in turn, cause local fluid flow through pore channels, resulting in viscous dissipation \cite{White1975, Johnson2001, Kobayashi2016, Gurevich2022}. Note that this behavior of attenuation is also consistent with the behavior of the bulk modulus during adsorption [Fig.~\ref{fig:Attenuation} (b)]: $K$ is almost constant until the liquid saturation reaches about 80\%, but then increases sharply to the value given by Gassmann's equation. In a remarkable similarity to macroporous media (\cite{Cadoret1998}, Fig.~1), this increase is quite sharp, but not instant. 

This behavior of both bulk modulus and attenuation is characteristic of near-uniform liquid saturation, which is attained when the characteristic linear scale of saturation heterogeneity is smaller than the so-called fluid diffusion length $l_{\rm d}=(K_{\rm f} \kappa / 2 \pi f \eta \phi)^{1/2}$, where $\kappa$ is the hydraulic permeability of the sample and $\eta$ the dynamic viscosity of the fluid, see~\cite{Gurevich2022}, Chapter~4 and references therein. Hydraulic permeability scales with the pore size squared, and for a carbonate rock corresponding to Figs.~1 and 2 of~\cite{Cadoret1998} is  on the order of $10^{-13}$~m$^2$, which for $f=10^3$~Hz gives $l_{\rm d}$ on the order of 1~cm. That is, the length scale of fluid saturation patches is below 1 cm.  In contrast, due its nano-porous structure, the Vycor glass has a permeability of $~5\cdot10^{-20}$~m$^2$~\cite{Vichit2000}, which for $f=5 \cdot 10^6$~Hz gives $l_{\rm d}$ of about 100~nm. The behavior of both bulk modulus and attenuation corresponding to near-uniform saturation indicates  that the water saturation  on adsorption is slightly heterogeneous on the length scale below 100~nm, or several pore sizes, but is uniform on larger scales (consistent with the fact that, on adsorption, the samples remain optically transparent for the entire range of vapor pressures). The sharp attenuation peak at the saturation of 90\% suggests that this attenuation only occurs when some portions of the pore space experience capillary condensation. This sharply increases the stiffness of these portions of the sample, and hence the contrast in elastic moduli between adjacent areas (some of which are still saturated only partially), which is essential for such attenuation to occur; see~\cite{Gurevich2022}, Chapters~3 and~4.

On desorption, both the bulk modulus and attenuation shown in Fig.~\ref{fig:Attenuation} exhibit much more gradual saturation dependence, which is typical of so-called patchy saturation, where the scale of saturation heterogeneity is larger than $l_{\rm d}$. This is consistent with the observed change in the optical properties of both samples during desorption when the hysteresis loop closed ($p/p_0 = 0.58$). At this point, the samples became white and opaque starting from the outer edge and spreading towards the center. The opacity reverted to translucency with the same radial progression between $p/p_0 = 0.58$ and $p/p_0 = 0.53$. This phenomenon was also observed by Page et al., who used light scattering measurements to show the formation of large-scale inhomogeneities in the pore network as the fluid drains. These inhomogeneities are large enough to scatter light, causing the samples to become opaque. The different saturation regions permit different ultrasound speeds, which arrive out of phase at the receiving transducer to affect the shape and apparent travel time of measured ultrasonic waves. This both contributes to attenuation and causes a dip in $v_{\rm L}$ just as the hysteresis loop is closing.

\subsection{Sample-Transducer Assembly}

\begin{figure}[h]
\centering
\includegraphics[width=0.7\linewidth]{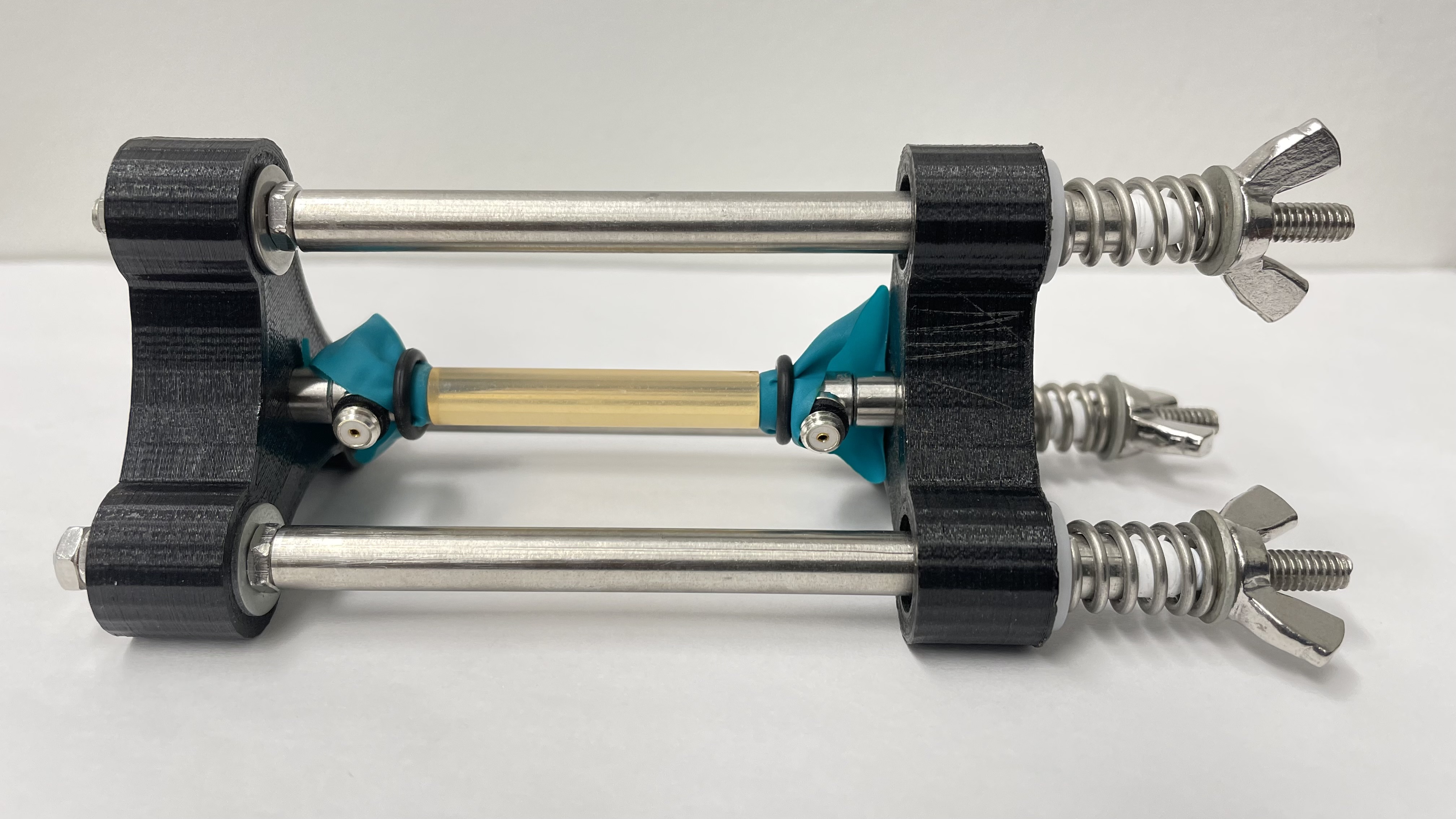}
\caption{Photo of the assembly used to measure the sound speed of the glass samples.}
\label{fig:Sample_Holder}
\end{figure}

Figure \ref{fig:Sample_Holder} shows the photo of the sample holder used in the experiments. The transducer surface is covered by a layer of nitrile, which acts as a couplant. The sample is held between the transducers by two 3D-printed plates of extruded polyethylene terephthalate glycol (PETG). The plate on the left is fixed in place. The plate on the right is movable. By rotating the wingnuts, pressure is applied to the sample via springs to ensure coupling across the transducer-sample interfaces.

{Between the lowest and the highest humidities considered in this work, we observed a 14\% increase in the mass of the nitrile couplant due to water absorption. Despite this, there was no effect on the travel time of ultrasonic waves through the couplant alone because its thickness was more than three times smaller than the wavelength ($\sim$ \SI{0.6}{mm}) of ultrasonic waves. Further, since the length of the couplant is 3 orders of magnitude less than that of the sample, any swelling upon humidification had negligible effect on the overall travel time of ultrasonic waves.}

\subsection{Potassium Acetate Desiccant}

The results listed in this work do not include the measurements where relative humidity was controlled by potassium acetate, used recently by Yurikov et al.~\cite{Yurikov2018} for studies of water adsorption on a sandstone sample. In solution, this salt hydrolyzes into acetic acid, which is volatile and readily releases into the air in the chamber, thus affecting the accuracy of the humidity sensor. It may also compete with water vapor for adsorption in nanopores, thus contaminating the sample. The expected RH of air above potassium acetate is 23 - 26\%. In practice, the reading on the RH probe approached the expected value (from room humidity: $\sim$30 - 40\%), before parabolically increasing to $\sim$75\%. After opening the chamber to ambient air, the reading increased to nearly 100\% and slowly returned to normal over the course of one hour. Potassium acetate was tested again without any samples in the chamber, producing the same result described above. After purging with air, the probe reported the correct humidities for all other salts. Subsequent ultrasound measurements with previous salts showed no difference before and after the use of potassium acetate, suggesting that any contamination was minimal or was undone by subsequent water adsorption and desorption. Because of the unreliability of the probe with this desiccant, and to prevent this effect from influencing water adsorption, potassium acetate was excluded from the experiments.

\subsection{Estimation of $K_{\rm s}$ and $G_{\rm s}$}

The bulk and shear moduli of the solid glass matrix ($K_{\rm s}$ and $G_{\rm s}$, respectively) were estimated using the modified Kuster and Toks\"oz (KT) theory. {The results for $K_{\rm s}$ are shown in Tables \ref{tab:Mechanical_Props}, ~\ref{tab:Vacuum_Table_V}, and ~\ref{tab:Vacuum_Table_C}}. Assuming that the samples have empty cylindrical inclusions of volume fraction $\phi$, the moduli of the solid phase can be calculated from the experimentally measured bulk and shear moduli of the dry samples ($K_0$ and $G_0$, respectively) using the following equations:
\begin{equation}
    (K_{\rm s} - K_0)\frac{K_{\rm s} + \frac{4}{3}G_{\rm s}}{K_0 + \frac{4}{3}G_{\rm s}} = \phi K_{\rm s} P^{\rm sp},
\label{eq:SI1}
\end{equation}
\begin{equation}
    (G_{\rm s} - G_0)\frac{G_{\rm s} + F_{\rm s}}{G_0 + F_{\rm s}} = \phi G_{\rm s} Q^{\rm sp},
\label{eq:SI2}
\end{equation}
where
\begin{equation}
    F_{\rm s} = \frac{G_{\rm s}}{6}\frac{9K_{\rm s} + 8G_{\rm s}}{K_{\rm s} + 2G_{\rm s}},
\label{eq:SI3}
\end{equation}
\begin{equation}
    P^{\rm sp} = \frac{K_{\rm s} + G_{\rm s}}{G_{\rm s}},
\label{eq:SI4}
\end{equation}
\begin{equation}
    P^{\rm sp} = \frac{1}{5}\left(\frac{16}{3} + 2\frac{G_{\rm s} + \gamma_{\rm s}}{\gamma_{\rm s}}\right),
\label{eq:SI5}
\end{equation}
and
\begin{equation}
    \gamma_{\rm s} = G_{\rm s}\frac{3K_{\rm s} + G_{\rm s}}{3K_{\rm s} + 7G_{\rm s}}.
\label{eq:SI6}
\end{equation}

\newpage


%

\end{document}